\pdfoutput=1
\documentclass[acmsmall,screen,nonacm]{acmart}

\usepackage{booktabs}   %
\usepackage{tikz}
\usepackage{algorithm}
\usepackage{algpseudocode}
\usepackage{float}
\usepackage{subfigure}
\usepackage{epstopdf} %
\usepackage{multirow} %
\usepackage{bm} %
\usepackage{xspace}
\usepackage{tcolorbox,enumitem}
\algnewcommand{\algorithmicmatch}{\textbf{match}}
\algnewcommand{\algorithmicwith}{\textbf{with}}
\algnewcommand{\algorithmiccase}{\textbf{case}}
\algblockdefx[MATCH]{Match}{EndMatch}
   [1]{\algorithmicmatch\ #1\ \algorithmicwith}
   [0]{}
\algnotext[MATCH]{EndMatch}
\algblockdefx[CASE]{Case}{EndCase}
   [1]{\algorithmiccase\ #1 :}
   [0]{}
\algnotext[CASE]{EndCase}
\algnotext{EndIf}
\algnotext{EndFor}
\algnotext{EndFunction}
\algnotext{EndWhile}
\algrenewcommand\algorithmicrequire{\textbf{Input:}}
\algrenewcommand\algorithmicensure{\textbf{Output:}}
\newcommand{\hide}[1]{}

\newcommand{\oa}[1]{\bar{#1}} %
\renewcommand{\vec}[1]{\bm{#1}} %
\newcommand{\nn}{N} %
\newcommand{\relu}{ReLU\xspace} %
\newcommand{\lb}{\langle} %
\newcommand{\rb}{\rangle} %
\newcommand{\ce}{\textit{cex}} %
\newcommand{\inc}{\ensuremath{\mathsf{inc}}\xspace} %
\newcommand{\dec}{\ensuremath{\mathsf{dec}}\xspace} %
\newcommand{\acasxu}{\ensuremath{\texttt{ACAS}\,\texttt{Xu}}\xspace}
\newcommand{\mnist}{\texttt{MNIST}\xspace}
\newcommand{\cifar}{\texttt{CIFAR-10}\xspace}

\newcommand{\myabstract}{\textsc{Abstract}\xspace}
\newcommand{\myabstractp}[1]{\textsc{Abstract}(#1)\xspace}
\newcommand{\refine}{\textsc{Refine}\xspace}
\newcommand{\refinep}[1]{\textsc{Refine}(#1)\xspace}
\newcommand{\verify}{\textsc{Verify}\xspace}
\newcommand{\verifyp}[1]{\textsc{Verify}(#1)\xspace}
\newcommand{\merge}{\textsc{Merge}\xspace}
\newcommand{\freeze}{\textsc{Freeze}\xspace}
\newcommand{\qfreeze}[1]{\ensuremath{\textsc{Freeze}^{#1}_{\texttt{+}}}\xspace} %
\newcommand{\absstep}{\texttt{abstract}\xspace}
\newcommand{\qabstract}[1]{\ensuremath{\texttt{q-abstract}^{#1}}\xspace} %
\newcommand{\propagate}{\textsc{Propagate}\xspace}
\newcommand{\mysplit}{\textsc{Split}\xspace}
\newcommand{\recover}{\textsc{Recover}\xspace}
\newcommand{\lbound}[2]{\mathop{\mathrm{lb}^{#1}}(#2)} %
\newcommand{\ubound}[2]{\mathop{\mathrm{ub}^{#1}}(#2)} %
\newcommand{\dg}{\mathcal{G}} %
\newcommand{\lossf}[1]{L^{\texttt{f}}(#1)}
\newcommand{\lossm}[1]{L^{\texttt{m}}(#1)}
\newcommand{\lossx}[1]{P_{\ce}(#1)}
\newcommand{\minloss}{\mathit{minLoss}}
\newcommand{\maxerr}{\mathit{maxGain}}
\newcommand{\bestop}{\mathit{bestStep}}

\newcommand{\vnncomp}{VNN-COMP'21}
\newcommand{\ourtool}{\textsc{NARv}\xspace} %
\newcommand{\marabou}{\textsc{Marabou}\xspace}
\newcommand{\planet}{\textsc{Planet}\xspace}
\newcommand{\ourtoolm}{\textsc{NARv[M]}\xspace} %
\newcommand{\ourtoolp}{\textsc{NARv[P]}\xspace} %
\newcommand{\aisqr}{\textsc{AI}$^2$\xspace}
\newcommand{\deepz}{\textsc{DeepZ}\xspace}
\newcommand{\deeppoly}{\textsc{DeepPoly}\xspace}
\newcommand{\reluval}{\textsc{ReluVal}\xspace}
\newcommand{\nnv}{\textsc{NNV}\xspace}
\newcommand{\deepsrgr}{\textsc{DeepSRGR}\xspace}
\newcommand{\cegarnn}{{\textsc{CEGAR-NN}}\xspace}
\newcommand{\cegarnni}{\mbox{\textsc{CEGAR-NN[I]}}\xspace}
\newcommand{\cegarnns}{\mbox{\textsc{CEGAR-NN[S]}}\xspace}
\newcommand{\tinc}[1]{$\vartriangle$ #1} %
\newcommand{\tdec}[1]{$\blacktriangledown$ #1} %
\newcommand{\ninc}[1]{$\blacktriangle$ #1} %
\newcommand{\x}{$\times$} %

\begin{document}

\title{Abstraction and Refinement: Towards Scalable and Exact Verification of Neural Networks}

\author{Jiaxiang Liu}
\email{jiaxiang0924@gmail.com}
\author{Yunhan Xing}
\email{xingyunhan@email.szu.edu.cn}
\author{Xiaomu Shi}
\email{xshi0811@gmail.com}
\affiliation{%
  \institution{Shenzhen University}
  \city{Shenzhen}
  \country{China}
}

\author{Fu Song}
\affiliation{%
  \institution{ShanghaiTech University}
  \city{Shanghai}
  \country{China}}
\email{songfu@shanghaitech.edu.cn}

\author{Zhiwu Xu}
\email{xuzhiwu@szu.edu.cn}
\author{Zhong Ming}
\affiliation{%
  \institution{Shenzhen University}
  \city{Shenzhen}
  \country{China}
}

\renewcommand{\shortauthors}{Liu et al.}

\begin{abstract}
As a new programming paradigm, deep neural networks (DNNs) have been
increasingly deployed in practice, but the lack of robustness
hinders their applications in safety-critical domains. While there are
techniques for verifying DNNs with formal guarantees, they are limited
in scalability and accuracy.
In this paper, we present a novel abstraction-refinement approach for scalable and exact DNN verification.
Specifically, we propose a novel abstraction  to break down the size of DNNs
by over-approximation.  The result of verifying the abstract DNN
is always conclusive if no spurious counterexample is reported.
To eliminate spurious counterexamples introduced by abstraction, we
propose a novel counterexample-guided refinement that refines the
abstract DNN to exclude a given spurious counterexample while still
over-approximating the original one.
 Our  approach is
orthogonal to and can be integrated with many existing verification
techniques.  For demonstration, we implement our approach using two
promising and exact tools \marabou and \planet as the underlying
verification engines, and evaluate on widely-used benchmarks \acasxu,
\mnist and \cifar. The results show that our approach can boost their
performance by solving more problems and reducing up to 86.3\% and 78.0\% verification time,
respectively. Compared to the most relevant abstraction-refinement
approach, our approach is 11.6--26.6 times faster. %
\end{abstract}

\maketitle

\section{Introduction}
Due to surprising breakthroughs in many challenging tasks such as
image recognition~\cite{DBLP:journals/ijcv/RussakovskyDSKS15} and natural
language processing~\cite{DBLP:journals/spm/X12a},
deep learning has arguably become a new programming paradigm that takes over traditional software programs in many
areas. For instance, deep neural networks (DNNs) are
increasingly being deployed in safety-critical
applications, e.g., autonomous
driving~\cite{DBLP:journals/expert/UrmsonW08} and medical
systems~\cite{DBLP:journals/mia/LitjensKBSCGLGS17}.
However, DNNs are fragile to small perturbations due to the lack of robustness~\cite{DalviDMSV04,SzegedyZSBEGF13,GoodfellowSS14,Carlini017,KurakinGB17a,PapernotMJFCS16}.
Therefore, it is important to formally guarantee the robustness of DNNs before
to deploy them in safety-critical applications.

Many efforts have been made to verify DNNs~\cite{DBLP:conf/cav/ElboherGK20,DBLP:conf/cav/PulinaT10,DBLP:conf/atva/Ehlers17,DBLP:conf/cav/KatzBDJK17,DBLP:conf/cav/KatzHIJLLSTWZDK19,DBLP:conf/cav/HuangKWW17,DBLP:conf/icml/WongK18,DBLP:conf/cvpr/LinYCZLLH19,DBLP:journals/corr/abs-1711-00455}.
Early work  relies
on using constraint solvers, often providing
soundness and completeness guarantees.
However, their scalability is  limited due to the intrinsic computational complexity, e.g.,
NP-complete even for simple neural networks and properties~\cite{DBLP:conf/cav/KatzBDJK17}.
Another line of work is based on abstract interpretation to improve the scalability at the cost of precision~\cite{DBLP:conf/sp/GehrMDTCV18,DBLP:conf/nips/SinghGMPV18,DBLP:journals/pacmpl/SinghGPV19,DBLP:conf/fm/TranLMYNXJ19,DBLP:conf/uss/WangPWYJ18,DBLP:conf/tacas/YangLLHWSXZ21}.
Although few of them incorporate refinement strategies to improve accuracy~\cite{DBLP:conf/uss/WangPWYJ18,DBLP:journals/pacmpl/SinghGPV19,DBLP:conf/tacas/YangLLHWSXZ21},
it remains a great challenge to efficiently and precisely verify large-scale DNNs.  %
One of the most promising techniques used in formal verification to improve
the efficiency is \emph{counterexample-guided abstraction refinement} (\emph{CEGAR}) framework~\cite{DBLP:conf/cav/ClarkeGJLV00}.
The essential idea of CEGAR is that, when given a target system $S$
to verify, an over-approximation, small-sized system $\oa{S}$ is constructed
by \emph{abstraction} and verified by an off-the-shelf tool.
The result is always conclusive
if no spurious counterexample is reported.
Otherwise, to regain precision, the abstract system $\oa{S}$
is refined guided by the spurious counterexample to exclude it.
The verification process is repeated on the refined system until the original system is proved
or a genuine counterexample is found.

To instantiate the CEGAR framework,
one needs to address the following two questions: (1) how to abstract
a target system and (2) how to refine an abstract system.
When instantiating CEGAR in DNN verification,
there are four technical challenges:
\begin{enumerate}%
\item[C1:]\label{i:ch1} The abstraction should guarantee
  \emph{soundness}, i.e., if an abstract DNN $\oa{\nn}$ is proved robust, the target
  DNN $\nn$ must be robust.
\item[C2:]\label{i:ch2} The abstraction should reduce the network
  size as much as possible while preserving accuracy as much as possible, %
   because coarse-grained abstract DNNs may %
  result in plenty of spurious counterexamples, thus requiring more refinement steps.
\item[C3:]\label{i:ch3} The refinement must preserve soundness as well, similar to C1, and also excludes
a given counterexample.
\item[C4:]\label{i:ch4} The refinement should regain the accuracy as
  much as possible, meanwhile enlarging the network size as little as
  possible.
\end{enumerate}

In this paper, addressing the above challenges,
we present a scalable and exact CEGAR-based approach for DNN verification
by proposing \emph{novel} procedures for abstraction and refinement.

We define the abstraction procedure as a synergistic integration of two novel \emph{abstraction primitives}: %
one is to merge neurons and the other is to remove neurons, both of which are able to reduce
network size. To address C1, these primitives are well-designed according to the weights and bounds of neurons,
thus provide soundness guarantees.
To address C2, the iterative application of abstraction primitives is guided by a
strategy that selects a
primitive aimed at minimizing the loss of accuracy during each iteration, thus can %
abstract out more neurons when sacrificing the same accuracy.

The refinement procedure is defined as a synergistic integration of two novel
\emph{refinement primitives}: one splits a single neuron into two neurons and the other recovers a
removed neuron. To address C3, the iterative application of refinement
primitives is restricted by a \emph{dependency graph}, a novel notion proposed to characterize
the dependency between refinement steps.
Under the restriction of the dependency graph, the refinement procedure is proved
sound, otherwise may not be sound.
Last but not least, the refinement procedure is also guided by
a strategy to address C4 which selects a refinement
primitive to regain the most accuracy, thus can keep the network size smaller when restoring the same amount of accuracy.

Our approach is orthogonal to and can be
integrated with many existing approaches.  For
evaluation, we implement our approach as a tool \ourtool using two
promising and exact tools
\marabou~\cite{DBLP:conf/cav/KatzHIJLLSTWZDK19} and
\planet~\cite{DBLP:conf/atva/Ehlers17} as back-end verification
engines. The experimental results show that our approach can improve
their scalability and boost their performance by reducing up to 86.3\% and 78.0\% verification
time, respectively. Moreover, our approach is illustrated to
significantly outperform the only tool that supports structure-oriented CEGAR-based
verification~\cite{DBLP:conf/cav/ElboherGK20}, 11.6--26.6 times faster.

To sum up, the main contributions of this work are as follows:
\begin{itemize}
  \item We propose a \emph{novel} abstraction procedure that synergistically integrates two abstraction primitives using a novel abstraction strategy,
  allowing to soundly and maximally reduce the network size when sacrificing the same accuracy.
  \item We propose a \emph{novel}  refinement procedure consisting of
  two refinement primitives, a notion of
  dependency graphs and a strategy for their synergistic integration,
   allowing to soundly refine the network and keep the
network size as small as possible when restoring the same amount of accuracy.
  \item We implement our approach as a tool \ourtool with two
    promising DNN verification engines \marabou and \planet and conduct an extensive evaluation,
demonstrating the efficacy of our approach.
\end{itemize}

\smallskip
\noindent
\textbf{Outline.}
Section~\ref{sec:prelim} defines basic notations. %
Section~\ref{sec:overview} presents the overview of our approach.
We propose our abstraction procedure and refinement procedure in Section~\ref{sec:abstraction} and Section~\ref{sec:refinement}, respectively.
Section~\ref{sec:experiment} reports experimental results.
Finally, after discussing related work in Section~\ref{sec:related},
we conclude the paper in Section~\ref{sec:conclusion}.

To foster further research, benchmarks and experimental data are
released at \url{https://github.com/NARv22/data}.  The source code
is available at \url{https://github.com/formes20/narv}.

\section{Preliminaries}\label{sec:prelim}

\subsection{Deep Neural Networks}

A \emph{fully connected feedforward deep neural network} (\emph{DNN}) with $\ell + 1$ layers is an acyclic
graph structured in layers, where
$0$-th and $\ell$-th layers are \emph{input layer} and
\emph{output layer}, respectively and the other layers are \emph{hidden layers}.
The nodes in each layer are \emph{neurons}.
 We use $v_{i,j}$ to denote the value of the $j$-th neuron in
layer $i$, and $\vec{v}_i = (v_{i,1},\ldots,v_{i,n})^T$ the output
vector of layer $i$ containing $n$ neurons. Sometimes, $v_{i,j}$ also
denotes the neuron itself.
Each neuron $v_{i,j}$ in layer $i$ ($1\leq i\leq \ell$) is associated with a
 \emph{bias} $b(v_{i,j})$, and is connected by the weighted edges $w(v_{i-1,k},v_{i,j})$ from the neurons $v_{i-1,k}$ in layer $i-1$.
A DNN computes the output of a given input by propagating it through the network, where the value of each neuron
is calculated by applying an
\emph{activation function} to the weighted sum of the neuron values from the preceding
layer.

Formally, a DNN is a function $\nn(\vec{x})$ defined
by:
\[\nn(\vec{x}) = W_{\ell} \vec{v}_{\ell-1} + \vec{b}_{\ell},\]
where $\vec{v}_0=\vec{x}$, $\vec{v}_i = \sigma(W_i \vec{v}_{i-1} + \vec{b}_i)$
for $1\le i < \ell$,
$W_{i}$ and $\vec{b}_i$ are respectively the weight matrix and bias
vector associated with layer $i$, and $\sigma$ is an activation
function applied in an element-wise manner. In this paper, we focus on the most
commonly used ReLU activation function $\relu(x)=\max(x,0)$.  %
The notation $\ubound{\nn}{v_{i,j}}$ (resp.,
$\lbound{\nn}{v_{i,j}}$) denotes an \emph{upper} (resp., \emph{lower})
\emph{bound} of $v_{i,j}$ in $\nn$, that is, $\lbound{\nn}{v_{i,j}}\le
v_{i,j} \le \ubound{\nn}{v_{i,j}}$, w.r.t. a given input space.

\subsection{Formal Verification of DNNs}
Given a DNN $\nn$, a property $P$ over the inputs $\vec{x}$
and a property $Q$ over outputs $\vec{y} = \nn(\vec{x})$, a
\emph{verification problem} $\varphi = \lb \nn,P,Q\rb$
is to check whether any input $\vec{x}$ that fulfils $P$ will result
in an output $\vec{y} = \nn(\vec{x})$ that satisfies $Q$, where
$P$ forms the input space of interests. As usual, we consider input properties in conjunctions of linear constraints.
W.l.o.g., we assume that the output layer only contains a single neuron $y$, and
the output property is of the form $y \le c$ for a given constant $c$~\cite{DBLP:conf/cav/ElboherGK20}.

Given a verification problem $\varphi = \lb \nn, P, Q\rb$,
a DNN $\oa{\nn}$ is an \emph{over-approximation}
of $\nn$ if $\nn(\vec{x})\le \oa{\nn}(\vec{x})$ for every
$\vec{x}$ that fulfils $P$. Note that $\nn(\vec{x})\le \oa{\nn}(\vec{x})$ implies that $\oa{\nn}(\vec{x}) \le c\Rightarrow\nn(\vec{x})\le c$.
An input $\vec{x}$ is a \emph{counterexample} of $\nn$ if
$\vec{x}$ fulfils $P$ but  $\nn(\vec{x})> c$.
A counterexample $\vec{x}$ of $\oa{\nn}$
is \emph{spurious} on $\nn$
if $\nn(\vec{x})\le c$.

\section{Overview of Our Approach}\label{sec:overview}

Our CEGAR-based approach
is described in Algorithm~\ref{alg:framework}, which
invokes two vital components: the abstraction
procedure \myabstract and the refinement procedure \refine.
Given a verification problem $\lb \nn, P, Q\rb$,
Algorithm~\ref{alg:framework} returns either YES indicating that
the problem holds or a counterexample $\ce$ as the witness of the violation.

\begin{algorithm}[t]
\caption{CEGAR-Based Framework of Our Approach}
\label{alg:framework}%
\begin{algorithmic}[1]
  \Require A verification problem $\lb \nn, P, Q \rb$
  \Ensure YES if the problem holds; otherwise a counterexample
  \State Build an abstract DNN $\oa{\nn} \gets
  \myabstractp{\nn,P}$ \label{l:framework:abstract}
  \While {\verifyp{$\lb \oa{\nn}, P, Q\rb$} = NO} \label{l:framework:verify}
    \State Extract a counterexample $\ce$
    \If {$\ce$ is a counterexample of $\lb \nn, P, Q \rb$}
      \State \Return $\ce$ \label{l:framework:real}
    \Else
        $~~\oa{\nn} \gets \refinep{\oa{\nn}, \ce}$\label{l:framework:refine}
    \EndIf
  \EndWhile
  \State \Return YES
\end{algorithmic}
\end{algorithm}

To solve a verification problem $\lb \nn, P, Q\rb$, we first build an over-approximation $\oa{\nn}$ of $\nn$ by invoking \myabstract
(line~\ref{l:framework:abstract}).
\myabstract first transforms $\nn$
into an equivalent DNN $\nn'$ such that increasing
the value of each single hidden neuron in $\nn'$ either increases or decreases
the network output, thus establishing monotonicity between the value of each hidden neuron and the network's output.
Then, we build the over-approximation $\oa{\nn}$ from
the DNN $\nn'$ by a synergistic integration
of two novel abstraction primitives: \merge and \freeze,
where \merge merges two neurons with the same monotonicity into a
single one while \freeze deletes a neuron from the
network. Both \merge and \freeze build an over-approximation of a given
DNN, thus provide soundness guarantees (i.e., challenge C1).
To address C2, abstraction primitives are iteratively applied according to
a strategy until a given accuracy threshold is reached,
where the strategy is designed to minimize the loss of accuracy during each iteration,
thus reduces the network size as much as possible when sacrificing the same accuracy.
To achieve this, we measure the loss of accuracy induced by
applying abstraction primitives. %

Next, we check if the verification problem
$\lb \oa{\nn},P,Q\rb$ holds or not by invoking
a verification engine \verify (line~\ref{l:framework:verify}).
If $\lb \oa{\nn},P,Q\rb$ holds, we can conclude that the original verification problem $\lb \nn,P,Q\rb$ holds as well,
because $\oa{\nn}$ over-approximates $\nn$.
Otherwise, a counterexample $\ce$ is extracted from $\oa{\nn}$.
If $\ce$ is a genuine counterexample of $\lb \nn,P,Q\rb$, $\ce$ is reported as the witness of the violation to the property $Q$ (line~\ref{l:framework:real}).
If $\ce$ is a spurious counterexample of $\lb \nn,P,Q\rb$,
$\oa{\nn}$ is refined to exclude $\ce$ by invoking \refine (line~\ref{l:framework:refine}).

\refine is also a synergistic integration
of two novel refinement primitives:  \mysplit and \recover,
where \mysplit splits an abstract neuron into two while \recover gets back a deleted
neuron. However, applying \mysplit or \recover without any restriction does not necessarily
yield an over-approximation of the original DNN $\nn$ (i.e., challenge C3).
To solve this issue, \mysplit and \recover are applied
according to a dependency graph, a novel notion proposed to characterize
their dependency.
To address C4, refinement primitives are iteratively applied
 according to a strategy until the spurious counterexample $\ce$ is excluded,
 where the strategy is designed to regain the most accuracy during each iteration, thus
 keeps the network size as small as possible when restoring the same amount of accuracy.
To achieve this, we introduce a profit function parameterized by the
counterexample $\ce$ to measure the accuracy that can be restored via
refinement primitives on different neurons.

\section{Network Abstraction}\label{sec:abstraction}

In this section, we present our abstraction procedure
\myabstract.

\subsection{Preprocessing}
\label{ss:preprocess}

Before to abstract a given DNN, %
all hidden neurons
should be classified into $\inc$ or $\dec$, indicating their monotonic
effect on the network's output. A neuron is $\inc$ if
increasing its value, while keeping all the inputs unchanged,
increases the network's output. Symmetrically, a neuron is $\dec$ if
decreasing its value increases the network's output.
However, not all hidden
neurons in an arbitrary DNN can be simply classified.  To achieve
this, we transform the given DNN $\nn$ into an equivalent DNN
$\nn'$, namely, $\nn(\vec{x})=\nn'(\vec{x})$  for any input $\vec{x}$,
during which the classification is performed on $\nn'$.

The preprocessing procedure starts with setting the
single output $y$ as $\inc$, and proceeds backwards layer by
layer. Suppose that neurons in layer $i+1$ have been classified as
$\inc$ or $\dec$. A hidden neuron $v_{i,j}$ in layer $i$ will be split
into two new neurons $v_{i,j}^+$ and $v_{i,j}^-$ by copying all
incoming edges of $v_{i,j}$. As for the outgoing edges, $v_{i,j}^+$
only keeps positive-weighted ones pointing from $v_{i,j}$ to $\inc$ neurons
and negative-weighted ones pointing from $v_{i,j}$ to $\dec$ neurons. Similarly,
$v_{i,j}^-$ only keeps positive-weighted ones\hide{pointing} from $v_{i,j}$ to $\dec$ neurons
and negative-weighted ones\hide{pointing} from $v_{i,j}$ to $\inc$ neurons. The new neurons
$v_{i,j}^+$ and $v_{i,j}^-$ are now respectively $\inc$ and $\dec$ neurons.

Formally, %
for every neuron $v_{i-1,k}$ in layer
$i-1$, we set:
\begin{align*}\small
{w}(v_{i-1,k},v_{i,j}^+) = {w}(v_{i-1,k},v_{i,j}^-) = w(v_{i-1,k},v_{i,j}).
\end{align*}
And for every neuron $v_{i+1,k}$ in layer $i+1$ of $\nn'$, we set:
\begin{align*}\small
w(v_{i,j}^+,v_{i+1,k}) &=
\left\{
  \begin{array}{ll}
    \!\!\!w(v_{i,j},v_{i+1,k}),
    &\!\!\!\!\!\mbox{if } w(v_{i,j},v_{i+1,k})\cdot s(v_{i+1,k}) > 0;\\
    \!\!\!0,
    &\!\!\!\!\!\mbox{otherwise}
  \end{array}
\right. \\
{w}(v_{i,j}^-,v_{i+1,k}) &=
\left\{
  \begin{array}{ll}
    \!\!\!w(v_{i,j},v_{i+1,k}),
    &\!\!\!\!\!\mbox{if } w(v_{i,j},v_{i+1,k})\cdot s(v_{i+1,k}) < 0;\\
    \!\!\!0,
    &\!\!\!\!\!\mbox{otherwise}
  \end{array}
\right.
\end{align*}
where $s(v)=1$ if $v$ is an \inc neuron and $s(v)=-1$  if $v$ is a \dec neuron.
Furthermore, the biases of new neurons are copied from $v_{i,j}$, i.e.,
$%
b(v_{i,j}^+) = b(v_{i,j}^-) = b(v_{i,j}).
$%

Intuitively, when the neuron $v_{i,j}^+$ increases, all the $\inc$ (resp.,
$\dec$) neurons in layer $i+1$ increase (resp., decrease) since they
connect to $v_{i,j}^+$ by positive (resp., negative) weights. Both the
increment of $\inc$ neurons and decrement of $\dec$ neurons in
layer $i+1$ will increase the network's output by definition. Hence
the neuron $v_{i,j}^+$ is $\inc$. The case of $v_{i,j}^-$ follows
similarly.

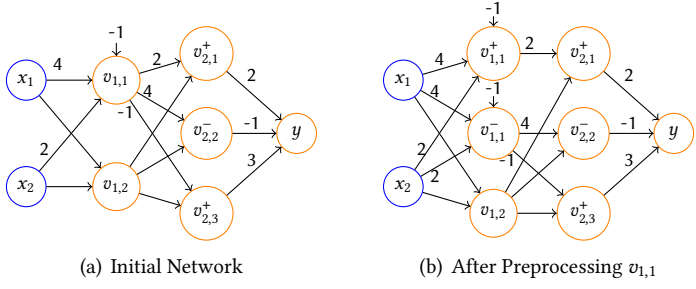
\begin{figure}[!t]
\centering\vspace{-2mm}%
\subfigure[Initial Network]{\label{fig:classify:init}
\begin{tikzpicture}[xscale=0.6, yscale=0.7]
  \tikzstyle{every node}=[scale=0.75, node font={\sf}]
  \tikzstyle{every circle node}=[minimum width=20pt, minimum
    height=20pt]
\node[circle,draw=blue] (x1) at(0,2){$x_1$};
\node[circle,draw=blue] (x2) at(0,0){$x_2$};
\node[circle,draw=orange] (v11) at(2,2){$v_{1,1}$};
\node[circle,draw=orange] (v12) at(2,0){$v_{1,2}$};
\node[circle,draw=orange] (v21) at(4,2.5){$v_{2,1}^+$};
\node[circle,draw=orange] (v22) at(4,1){$v_{2,2}^-$};
\node[circle,draw=orange] (v23) at(4,-0.5){$v_{2,3}^+$};
\node[circle,draw=orange] (y) at(6,1){$y$};
\draw[->] (x1) --(v11);
\draw[->] (x1) --(v12);
\draw[->] (x2) --(v11);
\draw[->] (x2) --(v12);
\draw[->] (v11) --(v21);
\draw[->] (v11) --(v22);
\draw[->] (v11) --(v23);
\draw[->] (v12) --(v21);
\draw[->] (v12) --(v22);
\draw[->] (v12) --(v23);
\draw[->] (v21) --(y);
\draw[->] (v22) --(y);
\draw[->] (v23) --(y);
\draw[->] (2,2.7) -- (v11);
\node at(2,2.9) {-1};
\node at(0.7,2.3) {4};
\node at(0.4,0.7) {2};
\node at(2.9,2.4) {2};
\node at(2.7,1.8) {4};
\node at(2.2,1.4) {-1};
\node at(5,2.1) {2};
\node at(5,1.2) {-1};
\node at(5,0.5) {3};
\end{tikzpicture}
}%
\qquad
\subfigure[After Preprocessing $v_{1,1}$]{\label{fig:classify:result}
\begin{tikzpicture}[xscale=0.6, yscale=0.7]
  \tikzstyle{every node}=[scale=0.75, node font={\sf}]
  \tikzstyle{every circle node}=[minimum width=20pt, minimum
    height=20pt]
\node[circle,draw=blue] (x1) at(0,2){$x_1$};
\node[circle,draw=blue] (x2) at(0,0){$x_2$};
\node[circle,draw=orange] (v11+) at(2,2.5){$v_{1,1}^+$};
\node[circle,draw=orange] (v11-) at(2,1){$v_{1,1}^-$};
\node[circle,draw=orange] (v12) at(2,-0.5){$v_{1,2}$};
\node[circle,draw=orange] (v21) at(4,2.5){$v_{2,1}^+$};
\node[circle,draw=orange] (v22) at(4,1){$v_{2,2}^-$};
\node[circle,draw=orange] (v23) at(4,-0.5){$v_{2,3}^+$};
\node[circle,draw=orange] (y) at(6,1){$y$};
\draw[->] (x1) --(v11+);
\draw[->] (x1) --(v11-);
\draw[->] (x1) --(v12);
\draw[->] (x2) --(v11+);
\draw[->] (x2) --(v11-);
\draw[->] (x2) --(v12);
\draw[->] (v11+) --(v21);
\draw[->] (v11-) --(v22);
\draw[->] (v11-) --(v23);
\draw[->] (v12) --(v21);
\draw[->] (v12) --(v22);
\draw[->] (v12) --(v23);
\draw[->] (v21) --(y);
\draw[->] (v22) --(y);
\draw[->] (v23) --(y);
\draw[->] (2,3.2) -- (v11+);
\draw[->] (2,1.7) -- (v11-);
\node at(2,3.35) {-1};
\node at(2,1.85) {-1};
\node at(0.8,2.4) {4};
\node at(0.7,1.8) {4};
\node at(0.4,0.7) {2};
\node at(0.7,0.2) {2};
\node at(2.8,2.7) {2};
\node at(2.7,1.2) {4};
\node at(2.3,0.5) {-1};
\node at(5,2.1) {2};
\node at(5,1.2) {-1};
\node at(5,0.5) {3};
\end{tikzpicture}
}\vspace{-2mm}
\caption{Example for Preprocessing}
\label{fig:classify}%
\end{figure}

\begin{example}
Consider the DNN shown in Figure~\ref{fig:classify:init}, where
the last two layers have been preprocessed. Consider neuron
$v_{1,1}$ whose bias is $-1$ and the related weights are shown.
The DNN after preprocessing $v_{1,1}$ is depicted in
Figure~\ref{fig:classify:result}, where:
$v_{1,1}$ is split into two neurons $v_{1,1}^+$ and $v_{1,1}^-$ that have the same incoming edges and
biases as $v_{1,1}$,
$v_{1,1}^+$ keeps the outgoing edge of weight $2$ pointing to the \inc neuron $v_{2,1}^+$,
$v_{1,1}^-$ keeps the outgoing edge of weight $4$ pointing to the \dec neuron $v_{2,2}^-$
and the outgoing edge of weight $-1$ pointing to the \inc neuron $v_{2,3}^+$.
The neurons $v_{1,1}^+$ and $v_{1,1}^-$ are now
respectively \inc and \dec.
\end{example}

After the preprocessing of all the hidden neurons, we obtain a
new DNN $\nn'$ that is equivalent to $\nn$, in which each hidden neuron is classified into either $\inc$ or $\dec$.
Thus,
we have:

\begin{lemma}
\label{lemma:transformation}
Any DNN $\nn$ can be transformed into an equivalent DNN $\nn'$ where
each hidden neuron is classified into either $\inc$ or $\dec$, by increasing
the network size by a factor of at most 2.
\end{lemma}

By Lemma~\ref{lemma:transformation}, we hereafter assume that
each given DNN has been preprocessed and all its hidden neurons have been classified into
\inc/\dec.

\subsection{Abstraction Primitives}
\label{ss:abstraction}

As aforementioned, we propose two novel abstraction primitives: \merge
and \freeze, to construct over-approximations of DNNs.

The \merge primitive is to merge a pair of hidden neurons with same
label \inc/\dec in the same layer into a single one. We seek to
increase the values of \inc neurons and decrease the values of \dec
neurons, ensuring that the network's output always increases.
Suppose we are constructing an over-approximation $\oa{\nn}$ of $\nn$.
Let $\oa{w}$ and $\oa{b}$ denote respectively the weights and biases
in the constructed network $\oa{\nn}$.
The \merge primitive merges two hidden \inc neurons $v_{i,j}$ and
$v_{i,k}$ into a new \inc neuron $v_{i,t}$ via the following steps:

\begin{enumerate}
  \item all edges connecting to $v_{i,j}$ or $v_{i,k}$ are removed;
  \item neurons $v_{i,j}$ and $v_{i,k}$ are replaced by a new neuron
    $v_{i,t}$;
  \item from each neuron $v_{i-1,p}$ in the preceding layer, an incoming
    edge to $v_{i,t}$ is added as %
    \[\oa{w}(v_{i-1,p},v_{i,t})=\max\{w(v_{i-1,p},v_{i,j}),\ w(v_{i-1,p},v_{i,k})\};\]
  \item to each neuron $v_{i+1,q}$ in the succeeding layer, an outgoing
    edge from $v_{i,t}$ is added as %
    \[\oa{w}(v_{i,t},v_{i+1,q})=w(v_{i,j},v_{i+1,q})+w(v_{i,k},v_{i+1,q});\]
  \item the bias of $v_{i,t}$ is
    $\oa{b}(v_{i,t})=\max\{b(v_{i,j}),\ b(v_{i,k})\}$.
\end{enumerate}

Intuitively, the $\max$ operation in steps (3) and (5) guarantees that
$v_{i,t}$ is no less than the original neurons
$v_{i,j}$ and $v_{i,k}$. By the definition of outgoing edges, this amounts to
increasing or keeping $v_{i,j}$ and $v_{i,k}$ in $\nn$.
Since $v_{i,j}$ and $v_{i,k}$ are both \inc, it ensures
that the output either does not change or is increased by \merge. Similarly,
the \merge primitive for \dec neurons is defined except that $\max$ is replaced
 by $\min$. A neuron produced by \merge is called
an \emph{abstract neuron}, otherwise an \emph{atomic neuron}.

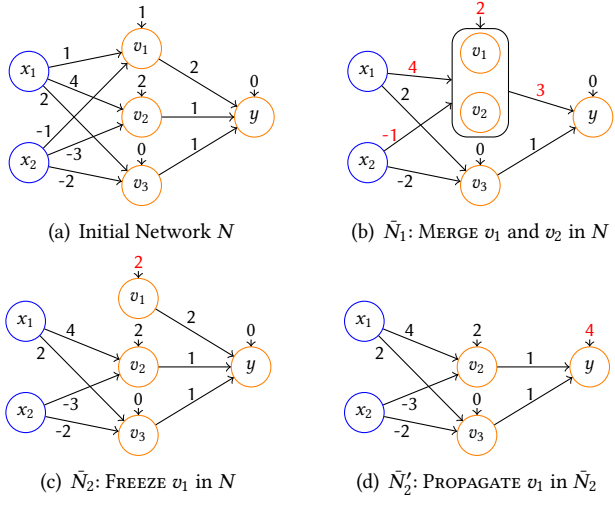
\begin{figure}[!t]
\centering
\subfigure[Initial Network $\nn$]{\label{fig:abstract:init}
\begin{tikzpicture}[yscale=0.6]
  \tikzstyle{every node}=[scale=0.75, node font={\sf}]
  \tikzstyle{every circle node}=[minimum width=20pt, minimum
    height=20pt]
\node[circle,draw=blue] (x1) at(0,2){$x_1$};
\node[circle,draw=blue] (x2) at(0,0){$x_2$};
\node[circle,draw=orange] (v1) at(1.5,2.5){$v_1$};
\node[circle,draw=orange] (v2) at(1.5,1){$v_2$};
\node[circle,draw=orange] (v3) at(1.5,-0.5){$v_3$};
\node[circle,draw=orange] (y) at(3,1){$y$};
\draw[->] (x1) --(v1);
\draw[->] (x1) --(v2);
\draw[->] (x1) --(v3);
\draw[->] (x2) --(v1);
\draw[->] (x2) --(v2);
\draw[->] (x2) --(v3);
\draw[->] (v1) --(y);
\draw[->] (v2) --(y);
\draw[->] (v3) --(y);
\draw[->] (1.5,3.1) -- (v1);
\draw[->] (1.5,1.6) -- (v2);
\draw[->] (1.5,0.1) -- (v3);
\draw[->] (3,1.6) -- (y);
\node at(1.5,3.3) {1};
\node at(1.5,1.8) {2};
\node at(1.5,0.3) {0};
\node at(3,1.8) {0};
\node at(0.5,2.4) {1};
\node at(0.6,1.8) {4};
\node at(0.2,1.4) {2};
\node at(0.2,0.7) {-1};
\node at(0.6,0.2) {-3};
\node at(0.5,-0.4) {-2};
\node at(2.2,2.1) {2};
\node at(2.2,1.2) {1};
\node at(2.2,0.4) {1};
\end{tikzpicture}
}
\qquad
\subfigure[$\oa{\nn}_1$: \merge $v_1$ and $v_2$ in $\nn$]{\label{fig:abstract:merge}
\begin{tikzpicture}[yscale=0.6]
  \tikzstyle{every node}=[scale=0.75, node font={\sf}]
  \tikzstyle{every circle node}=[minimum width=20pt, minimum
    height=20pt]
\node[circle,draw=blue] (x1) at(0,2){$x_1$};
\node[circle,draw=blue] (x2) at(0,0){$x_2$};
\node[circle,draw=orange] (v1) at(1.5,2.4){$v_1$};
\node[circle,draw=orange] (v2) at(1.5,1.1){$v_2$};
\node[circle,draw=orange] (v3) at(1.5,-0.5){$v_3$};
\node[circle,draw=orange] (y) at(3,1){$y$};
\node[rectangle,rounded corners=2mm ,draw=black ,minimum width=1cm, minimum height=1.9cm] (v1v2) at (1.5,1.75) {};
\draw[->] (x1) --(v1v2);
\draw[->] (x1) --(v3);
\draw[->] (x2) --(v1v2);
\draw[->] (x2) --(v3);
\draw[->] (v1v2) --(y);
\draw[->] (v3) --(y);
\draw[->] (1.5,3.2) -- (v1v2);
\draw[->] (1.5,0.1) -- (v3);
\draw[->] (3,1.6) -- (y);
\node [color=red] at(1.5,3.4) {2};
\node at(1.5,0.3) {0};
\node at(3,1.8) {0};
\node [color=red] at(0.6,2.1) {4};
\node at(0.5,1.5) {2};
\node [color=red] at(0.3,0.6) {-1};
\node at(0.5,-0.4) {-2};
\node [color=red] at(2.3,1.6) {3};
\node at(2.2,0.4) {1};
\end{tikzpicture}
}\\
\vspace{-3mm}
\subfigure[$\oa{\nn}_2$: \freeze $v_1$ in $\nn$]{\label{fig:abstract:freeze}
\begin{tikzpicture}[yscale=0.6]
  \tikzstyle{every node}=[scale=0.75, node font={\sf}]
  \tikzstyle{every circle node}=[minimum width=20pt, minimum
    height=20pt]
\node[circle,draw=blue] (x1) at(0,2){$x_1$};
\node[circle,draw=blue] (x2) at(0,0){$x_2$};
\node[circle,draw=orange] (v1) at(1.5,2.5){$v_1$};
\node[circle,draw=orange] (v2) at(1.5,1){$v_2$};
\node[circle,draw=orange] (v3) at(1.5,-0.5){$v_3$};
\node[circle,draw=orange] (y) at(3,1){$y$};
\draw[->] (x1) --(v2);
\draw[->] (x1) --(v3);
\draw[->] (x2) --(v2);
\draw[->] (x2) --(v3);
\draw[->] (v1) --(y);
\draw[->] (v2) --(y);
\draw[->] (v3) --(y);
\draw[->] (1.5,3.1) -- (v1);
\draw[->] (1.5,1.6) -- (v2);
\draw[->] (1.5,0.1) -- (v3);
\draw[->] (3,1.6) -- (y);
\node[color=red] at(1.5,3.3) {2};
\node at(1.5,1.8) {2};
\node at(1.5,0.3) {0};
\node at(3,1.8) {0};
\node at(0.6,1.8) {4};
\node at(0.2,1.3) {2};
\node at(0.6,0.2) {-3};
\node at(0.5,-0.4) {-2};
\node at(2.2,2.1) {2};
\node at(2.2,1.2) {1};
\node at(2.2,0.4) {1};
\end{tikzpicture}
}
\qquad
\subfigure[$\oa{\nn}'_2$: \propagate $v_1$ in $\oa{\nn}_2$]{\label{fig:abstract:propagate}
\begin{tikzpicture}[yscale=0.6]
  \tikzstyle{every node}=[scale=0.75, node font={\sf}]
  \tikzstyle{every circle node}=[minimum width=20pt, minimum
    height=20pt]
\node[circle,draw=blue] (x1) at(0,2){$x_1$};
\node[circle,draw=blue] (x2) at(0,0){$x_2$};
\node[circle,draw=orange] (v2) at(1.5,1){$v_2$};
\node[circle,draw=orange] (v3) at(1.5,-0.5){$v_3$};
\node[circle,draw=orange] (y) at(3,1){$y$};
\draw[->] (x1) --(v2);
\draw[->] (x1) --(v3);
\draw[->] (x2) --(v2);
\draw[->] (x2) --(v3);
\draw[->] (v2) --(y);
\draw[->] (v3) --(y);
\draw[->] (1.5,1.6) -- (v2);
\draw[->] (1.5,0.1) -- (v3);
\draw[->] (3,1.6) -- (y);
\node at(1.5,1.8) {2};
\node at(1.5,0.3) {0};
\node [color=red] at(3,1.8) {4};
\node at(0.6,1.8) {4};
\node at(0.2,1.3) {2};
\node at(0.6,0.2) {-3};
\node at(0.5,-0.4) {-2};
\node at(2.2,1.2) {1};
\node at(2.2,0.4) {1};
\end{tikzpicture}
}
\label{fig:abstract}
\vspace{-2mm}
\caption{Example for Abstraction Primitives}%
\end{figure}

\begin{example}
\label{ex:merge}
Consider the \inc neurons $v_1$ and $v_2$ of the DNN $\nn$ shown in
Figure~\ref{fig:abstract:init}. After merging $v_1$ and $v_2$,
we obtain the DNN $\oa{\nn}_1$ shown in
Figure~\ref{fig:abstract:merge}, where
for the abstract neuron of $(v_1,v_2)$,
the weight of its incoming edge from $x_1$ is $4 =\max\{1,4\}$,
its bias is $2 =\max\{1, 2\}$,
and the weight of its outgoing edge to $y$ is $3 = 2+1$.
Given an input $\vec{x_0} = (1,1)^T$, we
have $\oa{\nn}_1(\vec{x_0})=15 > 5 = \nn(\vec{x_0})$.
\end{example}

The following lemma justifies the soundness of \merge.

\begin{lemma}
\label{lemma:merge}
Let $\oa{\nn}$ be the DNN constructed from $\nn$ by a single
application of \merge. It holds that $\oa{\nn}(\vec{x}) \ge
\nn(\vec{x})$ for each input $\vec{x}$.
\end{lemma}

One application of \merge reduces the network size by 1, but may decrease the network's accuracy.
The induced inaccuracy sometimes
can be considerable, for instance, when the two weights in the $\max$
operation have a big difference as in Example~\ref{ex:merge}.
To avoid this issue, we introduce another abstraction
primitive \freeze, to freeze hidden neurons using constants.

Consider a hidden neuron $v_{i,j}$ and a constant $a$. If $a$ is an upper bound of $v_{i,j}$,
then freezing $v_{i,j}$ by $a$ amounts to increasing or keeping $v_{i,j}$.
Thus, the network's output is guaranteed to non-decrease when $v_{i,j}$ is \inc.
Similarly, if $a$ is a lower bound of the \dec neuron $v_{i,j}$,
the network's output is also guaranteed to non-decrease by applying \freeze. Formally, \freeze
constructs an over-approximation
$\oa{\nn}$ of $\nn$ as follows: for a hidden neuron $v_{i,j}$,
\begin{enumerate}
  \item all incoming edges to $v_{i,j}$ are removed, i.e.,
    \begin{displaymath}
      \oa{w}(v_{i-1,p},v_{i,j}) = 0,\
      \mbox{for each}\ v_{i-1,p}\ \mbox{in layer $i-1$;}
    \end{displaymath}
  \item the value of the neuron $v_{i,j}$ is replaced by a constant as
    \begin{displaymath}
      \oa{b}(v_{i,j}) =
        \left\{
          \begin{array}{ll}
            \ubound{\nn}{v_{i,j}}, & \mbox{ if } v_{i,j} \mbox{ is \inc};\\
            \lbound{\nn}{v_{i,j}}, & \mbox{ if } v_{i,j} \mbox{ is \dec.}
          \end{array}
        \right.
    \end{displaymath}
\end{enumerate}

Intuitively, \freeze produces neurons whose all incoming edges are weighted
by $0$ and their values are indeed their biases $\oa{b}(v_{i,j})$.
Neurons introduced by \freeze are called \emph{constant neurons}.

\begin{example}
\label{ex:freeze}
Assume that the range of the \inc neuron $v_1$ in
Figure~\ref{fig:abstract:init} is $[0,2]$. Then the network
$\oa{\nn}_2$ constructed from $\nn$ by \freeze on $v_1$ is shown in
Figure~\ref{fig:abstract:freeze}. All the incoming edges to $v_1$ are
removed.  Its bias is replaced with its upper bound $2$. Given the
input $\vec{x_0}=(1,1)^T$, $\oa{\nn}_2(\vec{x_0}) = 7 < 15 =
\oa{\nn}_1(\vec{x_0})$ shows that \freeze can be more accurate than
\merge in some situations.
\end{example}

The following lemma provides the soundness of \freeze.

\begin{lemma}
\label{lemma:freeze}
Let $\oa{\nn}$ be the DNN constructed from $\nn$ by a single
application of \freeze. It holds that $\oa{\nn}(\vec{x}) \ge
\nn(\vec{x})$ for each input $\vec{x}$.
\end{lemma}

One may notice that currently \freeze does not reduce the network size.
To eliminate constant neurons introduced by \freeze,
we propose a new procedure \propagate.
Consider a constant neuron $v_{i,j}$ in the DNN $\nn$,
\propagate works as follows:
\begin{enumerate}
  \item to propagate the value of $v_{i,j}$ to the succeeding layer $i+1$,
    for each neuron $v_{i+1,q}$ in layer $i+1$, we set:
    \begin{displaymath}
      b'(v_{i+1,q}) = b(v_{i+1,q}) + w(v_{i,j},v_{i+1,q})\cdot b(v_{i,j});
    \end{displaymath}
  \item the constant neuron $v_{i,j}$ and related edges are removed.
\end{enumerate}

\begin{example}
\label{ex:propagate}
Consider the constant neuron $v_1$ in the DNN $\oa{\nn}_2$ shown in
Figure~\ref{fig:abstract:freeze}.
By applying \propagate to $v_1$, $v_1$ is removed and its value $2$
is propagated to the neuron $y$, resulting in the DNN
shown in Figure~\ref{fig:abstract:propagate}.
\end{example}

By definition, the obtained DNN $\nn'$ after \propagate is
equivalent to $\nn$. Thus, we get:

\begin{lemma}
\label{lemma:propagate}
Let $\nn'$ be the DNN constructed from $\nn$ by a single
application of \propagate on a constant neuron. It holds that $\nn'(\vec{x}) =
\nn(\vec{x})$ for each input $\vec{x}$.
\end{lemma}

 \freeze and \propagate can cooperate to delete
hidden neurons, hence reducing  network size.
We design \propagate as an individual procedure instead of
merging with \freeze due to the following reasons:
(i) to improve abstraction efficiency, \propagate is invoked only once
before invoking the verification engine \verify;
(ii) to keep \freeze \emph{local}, i.e., without affecting the succeeding layer,
as locality makes abstraction steps less dependent on
each other, thus allows us to extend abstraction primitives further in
Section~\ref{ss:pre-abstraction}.

Following Lemmas~\ref{lemma:merge},~\ref{lemma:freeze} and~\ref{lemma:propagate}, we
conclude that our two abstraction primitives \merge and \freeze (followed by \propagate) do construct
over-approximations.

\begin{corollary}
\label{col:abstraction}
Let $\oa{\nn}$ be the DNN obtained from $\nn$ by iteratively applying
abstraction primitives: \merge and/or \freeze (followed by \propagate).  It holds that
$\oa{\nn}(\vec{x}) \ge \nn(\vec{x})$ for each input $\vec{x}$.
\end{corollary}

Given a DNN $\nn$ to verify, after iteratively applying our
abstraction primitives, we get an abstract DNN $\oa{\nn}$.
Corollary~\ref{col:abstraction} ensures that $\oa{\nn}$ is an
over-approximation of $\nn$, namely, if the specified
property holds for $\oa{\nn}$, it holds as well for $\nn$.
Therefore, we solve the challenge C1.

\subsection{Generalizing the \freeze Primitive}
\label{ss:pre-abstraction}

Starting from a DNN $\nn_0$, iterating the application of abstraction
primitives derives a sequence of DNNs $\nn_1,\nn_2,\cdots,\nn_k$.
Corollary~\ref{col:abstraction} guarantees that $\nn_j$ is an
over-approximation of $\nn_i$ for any $0\le i\le j\le k$.
Recall that when
applying \freeze on some neuron in $\nn_i$, its bounds with respect to
current DNN $\nn_i$ is required by definition. Since the structures of DNNs change along the sequence of DNNs due to abstraction, the bounds in
$\nn_i$ may be different from those in $\nn_j$ if $i\not=j$. To apply
the abstraction primitive \freeze, a naive approach is to calculate the bounds each
time before applying \freeze. The calculation is no doubt a considerable
overhead.
To mitigate this issue, we generalize the abstraction primitive \freeze
based on the following observation.

Observe that an abstraction primitive applied to layer $i$
does not change the values of neurons in layers $j<i$, thus their bounds.
A more efficient way is to calculate the bounds only once in the
initial network $\nn_0$, and all abstraction primitives are applied
backwards layer by layer. That is, an abstraction primitive to layer $i$
must be applied \emph{after} the applications of primitives to layers
$j> i$. This constrained order enables all \freeze primitives to
make use of the bounds calculated in the DNN $\nn_0$, thus avoids
the considerable calculation overhead, at the cost of
flexibility when performing abstraction.
However, it is sometimes too
restricted to achieve a good abstraction.
To gain more flexibility
while preserving the efficiency, we generalize the primitive \freeze
as follows.

A generalized primitive $\qfreeze{M}$ of \freeze is defined to
freeze a neuron w.r.t. a given DNN $M$. We emphasize that
$\qfreeze{M}$ is parameterized by the given DNN $M$.
To construct a network $\oa{\nn}$ from $\nn$ by applying
$\qfreeze{M}$ on a hidden neuron $v_{i,j}$ of $\nn$,
$\qfreeze{M}$ works almost the same as \freeze except
that $\qfreeze{M}$ leverages bounds w.r.t. the given DNN $M$
instead of $\nn$.
More specifically, $\qfreeze{M}$ works as follows:
\begin{enumerate}
  \item all incoming edges to $v_{i,j}$ are removed, i.e.,
    \begin{displaymath}
      \oa{w}(v_{i-1,p},v_{i,j}) = 0,\
      \mbox{for each}\ v_{i-1,p}\ \mbox{in layer $i-1$;}
    \end{displaymath}
  \item the value of the neuron $v_{i,j}$ is replaced by a constant as
    \begin{displaymath}
      \oa{b}(v_{i,j}) =
        \left\{
          \begin{array}{ll}
            \ubound{M}{v_{i,j}}, & \mbox{ if } v_{i,j} \mbox{ is \inc};\\
            \lbound{M}{v_{i,j}}, & \mbox{ if } v_{i,j} \mbox{ is \dec.}
          \end{array}
        \right.
    \end{displaymath}
\end{enumerate}

Compared to \freeze, the bounds $\ubound{M}{v_{i,j}}$ and
$\lbound{M}{v_{i,j}}$ are used instead of $\ubound{\nn}{v_{i,j}}$ and
$\lbound{\nn}{v_{i,j}}$.

As a result, Lemma~\ref{lemma:freeze} cannot be generalized
to the new primitive $\qfreeze{M}$.
That is, the constructed $\oa{\nn}$ does not necessarily over-approximate $\nn$.
Nevertheless, we can show that Lemma~\ref{lemma:freeze} can be generalized
to $\qfreeze{M}$ if
$\qfreeze{M}$ conforms to the definition of \freeze, i.e.,
$\ubound{M}{v_{i,j}}$ (resp., $\lbound{M}{v_{i,j}}$) is also an upper
(resp., lower) bound of $v_{i,j}$ in $\nn$.
Thus, we have:

\begin{lemma}
\label{lemma:qfreeze}
Given two DNNs $\nn$ and $M$, let $\oa{\nn}$ be the constructed
network from $\nn$ by $\qfreeze{M}$ on neuron $v$. $\oa{\nn}(\vec{x})
\ge \nn(\vec{x})$ holds for each input $\vec{x}$, if the upper bound
$\ubound{M}{v}$ (resp., lower bound $\lbound{M}{v}$) is also an upper
(resp., lower) bound of $v$ in $\nn$.
\end{lemma}

Given a DNN $M$, $\qfreeze{M}$ is called a \emph{quasi-abstraction
  primitive w.r.t. $M$}, where the prefix "quasi" indicates that it is not
yet ready to be used as an abstraction primitive, since it does not
have a strong property like Lemma~\ref{lemma:freeze}.

To leverage $\qfreeze{M}$ in the abstraction procedure, we define an
\emph{abstraction step} as the instance of applying an abstraction
primitive on one or two specific neurons.  Abstraction steps thus
relate abstraction primitives to their target neurons. We denote
abstraction steps by \absstep steps. Similarly, we have \merge,
\freeze and $\qfreeze{M}$ steps. We refer to \merge and $\qfreeze{M}$
steps together as \emph{quasi-abstraction steps w.r.t. $M$}, denoted
by $\qabstract{M}$ steps. The \qabstract{M} steps have the following
important property:

\begin{lemma}
\label{lemma:permutation}
Given two DNNs $\nn$ and $M$, %
let $\oa{\nn}$ be the network constructed
from $\nn$ via iteratively applying the sequence $\tau$ of $\qabstract{M}$ steps.
A \emph{permutation} $\tau'$
of $\tau$ can be constructed such that
\begin{enumerate}
  \item [(i)] each $\qfreeze{M}$ step precedes all \merge steps in
    $\tau'$;
  \item [(ii)] all $\qfreeze{M}$ steps in $\tau'$ are applied
    backwards layer by layer;
  \item [(iii)] the subsequence of \merge steps in $\tau'$ is
    identical to that in $\tau$;
  \item [(iv)] the network $\oa{\nn}'$ obtained by applying $\tau'$
    on $\nn$ is identical to the network $\oa{\nn}$.
\end{enumerate}
We call $\tau'$ an \emph{implicit order} or \emph{implicit sequence}
of $\tau$.
\end{lemma}

Lemma~\ref{lemma:permutation} indicates that a series of \qabstract{M}
steps can be reordered, such that all \qfreeze{M} steps are firstly
performed backwards layer by layer, and then the \merge steps are
performed. Moreover, the reordering does not change the resulting
network. Selecting the initial target DNN $\nn$ as $M$ used for the generalized
primitive \qfreeze{M}, each $\qfreeze{\nn}$ step in the implicit
sequence builds an over-approximation by
Lemma~\ref{lemma:qfreeze}. The following theorem justifies that
$\qabstract{\nn}$ steps, therefore \qfreeze{\nn}, can be leveraged to
perform abstraction:

\begin{theorem}
\label{thm:qabstract}
Let $\oa{\nn}$ be the DNN constructed from $\nn$ by a series of
$\qabstract{\nn}$ steps. It holds that $\oa{\nn}(\vec{x}) \ge
\nn(\vec{x})$ for each input $\vec{x}$.
\end{theorem}

To abstract a given DNN $\nn$, Theorem~\ref{thm:qabstract} guarantees
that we only need to calculate the bounds of neurons once on the initial
$\nn$. The quasi-abstraction primitive $\qfreeze{\nn}$ can be used for
abstraction instead of \freeze. In the abstraction procedure,
primitives \merge and \qfreeze{\nn} can be applied to any neuron in any
order
to construct over-approximations. The soundness still holds by
Theorem~\ref{thm:qabstract}.

In the rest of paper, since we will always use the initial network
$\nn$ for the generalized primitive \qfreeze{\nn} and the calculation
of bounds, for the sake of simplicity, we use the notation \qfreeze{}
for \qfreeze{\nn}. And when referring to \emph{abstraction
steps/primitives}, we now mean \merge and \qfreeze{}, thanks to
Theorem~\ref{thm:qabstract}.

\subsection{Abstraction Strategy}
\label{ss:guided-abstraction}

Having only abstraction primitives is not enough to accomplish the
abstraction procedure due to challenge C2, i.e., reducing the network
size as much as possible while preserving accuracy as much as
possible.  To build an initial abstraction addressing challenge C2,
two questions have to be answered: \emph{what should be done for a
single abstraction step} and \emph{how many steps should be
performed?}

The objective for the first question is to introduce less inaccuracy
at each abstraction step, thus admitting more abstraction steps and
reducing the size more, when sacrificing the same accuracy. For this
purpose, we propose to first locate a less important neuron and then
abstract it, where a neuron is less important if it contributes less
to the network's output. We measure the importance of neurons by their
values, where the value $V(v_{i,j})$ of a hidden neuron $v_{i,j}$ is
estimated using the mean of its upper and lower bounds, i.e.,
\[V(v_{i,j}) = \frac{1}{2}(\ubound{}{v_{i,j}} + \lbound{}{v_{i,j}}).\]
After a neuron
$v_{i,j}$ with the minimum estimated value $V(v_{i,j})$ is found, we
need decide which abstraction primitive--\merge or \qfreeze{}--should be applied to minimize the loss
of accuracy. Therefore, we measure the loss of accuracy
induced by abstraction primitives, based on which abstraction primitive is chosen.

The loss $\lossf{v_{i,j}}$ of accuracy by applying \qfreeze{} on a neuron
$v_{i,j}$ is measured by the difference
between the estimated value $V(v_{i,j})$ of $v_{i,j}$ and the constant $\oa{b}(v_{i,j})$ used for freezing $v_{i,j}$: %
\begin{displaymath}
  \lossf{v_{i,j}} = |\oa{b}(v_{i,j}) - V(v_{i,j})|.
\end{displaymath}

As applying \merge on two neurons $v_{i,j}$ and $v_{i,k}$ produces an abstract
neuron $v_{i,t}$, the loss $\lossm{v_{i,j},v_{i,k}}$ of accuracy
by applying \merge on $v_{i,j}$ and $v_{i,k}$
should depend on the changes on weights as
well as the values of both neurons.
Thus, a linear combination of the
estimated values is used to estimate $\lossm{v_{i,j},v_{i,k}}$:
\begin{displaymath}
  \lossm{v_{i,j},v_{i,k}} = R(v_{i,j},v_{i,t}) \cdot V(v_{i,j}) +
  R(v_{i,k},v_{i,t}) \cdot V(v_{i,k})
\end{displaymath}
where each coefficient $R(\cdot)$ is a ratio characterizing changes on
the weights of incoming edges to $v_{i,t}$:
\begin{displaymath}
  R(v_{i,j},v_{i,t}) = \frac{\sum_p|\oa{w}(v_{i-1,p},v_{i,t}) -
    w(v_{i-1,p},v_{i,j})|}{\sum_p |w(v_{i-1,p},v_{i,j})|}.
\end{displaymath}

Based on the above loss measurements, we propose a value-guided abstraction
strategy, described in Algorithm~\ref{alg:abstraction},
to synergistically apply abstraction primitives. It determines which
abstraction step should be chosen at the current iteration.
It starts by selecting the neuron with the minimum estimated value $V(v_{i,j})$ as
the target neuron (line~\ref{l:abs:strategy:min}). Then it chooses
among all available abstraction steps the one that loses the least
accuracy %
(lines~\ref{l:abs:strategy:loss:begin}--\ref{l:abs:strategy:loss:end}).
At the end, this optimum abstraction step is applied to the network
$\nn$ resulting in an over-approximation $\oa{\nn}$
(line~\ref{l:abs:strategy:apply}).
Furthermore, the mapping $V(\cdot)$
is updated accordingly for next use.

\begin{algorithm}[t]
\caption{Value-Guided Abstraction Strategy}
\label{alg:abstraction}
\begin{algorithmic}[1]
  \Require A DNN $N$, a mapping $V$ estimating values
  \Ensure An over-approximation $\oa{\nn}$ of $N$, updated $V$
    \State Select $v_{i,j}$ with the minimum estimated value $V(v_{i,j})$\label{l:abs:strategy:min}
    \State $\minloss \gets \infty$, $\bestop \gets \bot$
    \For {each hidden neuron $v_{i,k}$ of same label as $v_{i,j}$}\label{l:abs:strategy:loss:begin}
      \If {$\lossm{v_{i,j},v_{i,k}} < \minloss$}
        \State $\!\minloss \gets \lossm{v_{i,j},v_{i,k}}$, $\bestop \gets \merge(v_{i,j},v_{i,k})$
      \EndIf
    \EndFor
    \If {$\lossf{v_{i,j}} < \minloss$}
      \State $\minloss \gets \lossf{v_{i,j}}$, $\bestop\gets\qfreeze{}(v_{i,j})$\label{l:abs:strategy:loss:end}
    \EndIf
    \State Apply $\bestop$ on $\nn$ to construct $\oa{\nn}$\label{l:abs:strategy:apply}
    \State Update $V$ according to $\bestop$
    \State \Return $\oa{\nn}$, $V$
\end{algorithmic}
\end{algorithm}

For the second question in generating the initial abstraction, we
adopt the terminating condition proposed
in~\cite{DBLP:conf/cav/ElboherGK20}. A set $X$ of inputs satisfying
the input property $P$ is sampled before applying abstraction.
The \myabstract procedure iteratively
applies abstraction steps to build a sequence of over-approximations $\nn_1,\nn_2,\cdots,\nn_k$ according to the value-guided abstraction strategy (i.e., Algorithm~\ref{alg:abstraction})
until $\nn_k(\vec{x})$ for some input $\vec{x}\in X$ violates the output property $Q$.

Remark that the value-guided abstraction strategy together with the terminating
condition is a trade-off for the challenge C2.

\section{Network Refinement}\label{sec:refinement}
In this section, we present our refinement procedure
\refine.

\subsection{Refinement Primitives}
\label{ss:refinement}
Given a verification problem $\lb \nn, P, Q\rb$ and
a counterexample $\ce$ of $\oa{\nn}$ which is spurious on $\nn$,
to address challenge C3, the goal of refining an over-approximation $\oa{\nn}$ of the DNN $\nn$
is defined to construct a network $\oa{\nn}'$ satisfying the following condition:
\begin{equation*}\label{eqn:refine}
\oa{\nn}(\vec{x}) \ge \oa{\nn}'(\vec{x}) \ge \nn(\vec{x})
\mbox{ for each } \vec{x}  \mbox{ and } \oa{\nn}'(\ce) \mbox{ satisfies } Q.
\end{equation*}
That means refinement primitives should construct a DNN
$\oa{\nn}'$ over-approximating $\nn$, while $\oa{\nn}'$ is itself
over-approximated by $\oa{\nn}$ to exclude the spurious counterexample $\ce$ of $\nn$.

We define refinement primitives as the inverses of abstraction
primitives. Particularly, \emph{refinement steps}, the instances of
refinement primitives, are the inverses of abstraction steps.
Therefore, we introduce two refinement primitives: \mysplit and \recover,
corresponding to the abstraction primitives \merge and \qfreeze{}, respectively.
The \mysplit primitive splits an abstract neuron back into
two neurons that were merged by \merge during abstraction and the \recover primitive recovers a
constant neuron into the status before \qfreeze{}.

Now the question is, \emph{can we freely choose any
abstract or constant neuron in $\oa{\nn}$ to apply a refinement step without damaging soundness (i.e., challenge C3), like what we
do when applying an abstraction step?} Unfortunately, the answer is negative.

\begin{figure}[!t]
\centering\vspace{-2mm}
\begin{tikzpicture}[xscale=0.9, yscale=0.4]
  \tikzstyle{every node}=[scale=0.75, node font={\sf}]
  \tikzstyle{every circle node}=[minimum width=20pt, minimum
    height=20pt]
\begin{scope}%
\node at(1.75,-1) {{\large ($\nn_0$)}};
\node[circle,draw=blue] (x1) at(0,1){$x_1$};
\node[circle,draw=orange] (v11) at(1,2){$v_1$};
\node[circle,draw=orange] (v12) at(1,0){$v_2$};
\node[circle,draw=orange] (v13) at(2.5,2){$v_3$};
\node[circle,draw=orange] (v14) at(2.5,0){$v_4$};
\node[circle,draw=orange] (y1) at(3.5,1){$y$};
\draw[->] (x1) --(v11);
\draw[->] (x1) --(v12);
\draw[->] (v11) --(v13);
\draw[->] (v12) --(v13);
\draw[->] (v11) --(v14);
\draw[->] (v12) --(v14);
\draw[->] (v13) --(y1);
\draw[->] (v14) --(y1);

\node at(0.4,1.7) {1};
\node at(0.4,0.3) {1};
\node at(1.6,2.3) {3};
\node at(1.5,1.6) {1};
\node at(1.4,0.8) {1};
\node at(1.6,0.3) {2};
\node at(3.1,1.7) {1};
\node at(2.9,0.7) {1};
\end{scope}

\path[-stealth, line width=1.5pt] (0,0) edge (-0.8,-1);
\path[-stealth, line width=1.5pt] (3.5,0) edge (4.3,-1);
\node at(-1.5,-0.3) {$\merge(v_3,v_4)$};
\node at(4.9,-0.3) {$\merge(v_1,v_2)$};

\begin{scope}[yshift=-15]%
\node at(-0.8,-4.5) {{\large ($\oa{\nn_1}$)}};
\node[circle,draw=blue] (x2) at(-2.5,-2.5){$x_1$};
\node[circle,draw=orange] (v21) at(-1.5,-1.5){$v_1$};
\node[circle,draw=orange] (v22) at(-1.5,-3.5){$v_2$};
\node[circle,draw=orange] (v23) at(0,-1.5){$v_3$};
\node[circle,draw=orange] (v24) at(0,-3.5){$v_4$};
\node[circle,draw=orange] (y2) at(1,-2.5){$y$};
\node at(-2.1,-1.8) {1};
\node at(-2.1,-3.2) {1};
\node at(-0.8,-1.6) {3};
\node at(-0.8,-3.3) {2};
\node at(0.55,-2.1) {2};
\node[rectangle,rounded corners=1mm ,draw=black ,minimum width=0.9cm, minimum height=2cm] (2v3v4) at (0,-2.5) {};
\draw[->] (x2) --(v21);
\draw[->] (x2) --(v22);
\draw[->] (v21) --(2v3v4);
\draw[->] (v22) --(2v3v4);
\draw[->] (2v3v4) --(y2);
\end{scope}

\begin{scope}[yshift=-15]%
\node at(4.3,-4.5) {{\large ($\oa{\nn_3}$)}};
\node[circle,draw=blue] (x3) at(2.5,-2.5){$x_1$};
\node[circle,draw=orange] (v31) at(3.5,-1.5){$v_1$};
\node[circle,draw=orange] (v32) at(3.5,-3.5){$v_2$};
\node[circle,draw=orange] (v33) at(5,-1.5){$v_3$};
\node[circle,draw=orange] (v34) at(5,-3.5){$v_4$};
\node[circle,draw=orange] (y3) at(6,-2.5){$y$};
\node[rectangle,rounded corners=1mm ,draw=black ,minimum width=0.9cm, minimum height=2cm] (3v1v2) at (3.5,-2.5) {};
\node at(2.9,-2.1) {1};
\node at(4.2,-1.6) {4};
\node at(4.2,-3.3) {3};
\node at(5.5,-1.7) {1};
\node at(5.5,-3.3) {1};
\draw[->] (x3) --(3v1v2);
\draw[->] (3v1v2) --(v33);
\draw[->] (3v1v2) --(v34);
\draw[->] (v33) --(y3);
\draw[->] (v34) --(y3);
\end{scope}

\path[-stealth, line width=1.5pt] (-0.8,-5.6) edge (-0.8,-6.6);
\path[-stealth, line width=1.5pt] (4.3,-5.6) edge (4.3,-6.6);
\node at(0.2,-6) {$\merge(v_1,v_2)$};
\node at(5.3,-6) {$\merge(v_3,v_4)$};

\begin{scope}[yshift=-45]%
\node at(-0.8,-9.2) {{\large ($\oa{\nn_2}$)}};
\node[circle,draw=blue] (x4) at(-2.5,-7){$x_1$};
\node[circle,draw=orange] (v41) at(-1.5,-6){$v_1$};
\node[circle,draw=orange] (v42) at(-1.5,-8){$v_2$};
\node[circle,draw=orange] (v43) at(0,-6){$v_3$};
\node[circle,draw=orange] (v44) at(0,-8){$v_4$};
\node[circle,draw=orange] (y4) at(1,-7){$y$};

\node[rectangle,rounded corners=1mm ,draw=black ,minimum width=0.9cm, minimum height=2cm] (4v3v4) at (0,-7) {};
\node[rectangle,rounded corners=1mm ,draw=black ,minimum width=0.9cm, minimum height=2cm] (4v1v2) at (-1.5,-7) {};

\draw[->] (x4) --(4v1v2);
\draw[->] (4v1v2) --(4v3v4);
\draw[->] (4v3v4) --(y4);

\node at(-2.1,-6.6) {1};
\node [color=red] at(-0.8,-6.6) {\textbf{5}};
\node at(0.55,-6.6) {2};
\end{scope}

\begin{scope}[yshift=-45]%
\node at(4.3,-9.2) {{\large ($\oa{\nn_4}$)}};
\node[circle,draw=blue] (x5) at(2.5,-7){$x_1$};
\node[circle,draw=orange] (v51) at(3.5,-6){$v_1$};
\node[circle,draw=orange] (v52) at(3.5,-8){$v_2$};
\node[circle,draw=orange] (v53) at(5,-6){$v_3$};
\node[circle,draw=orange] (v54) at(5,-8){$v_4$};
\node[circle,draw=orange] (y5) at(6,-7){$y$};

\node[rectangle,rounded corners=1mm ,draw=black ,minimum width=0.9cm, minimum height=2cm] (5v1v2) at (3.5,-7) {};
\node[rectangle,rounded corners=1mm ,draw=black ,minimum width=0.9cm, minimum height=2cm] (5v3v4) at (5,-7) {};

\draw[->] (x5) --(5v1v2);
\draw[->] (5v1v2) --(5v3v4);
\draw[->] (5v3v4) --(y5);

\node at(2.9,-6.6) {1};
\node [color=red] at(4.2,-6.6) {\textbf{4}};
\node at(5.55,-6.6) {2};
\end{scope}
\end{tikzpicture}
\vspace*{-2mm}
\caption{Different Orders of Applying Same \merge Steps}
\label{fig:abstract-order}%
\end{figure}
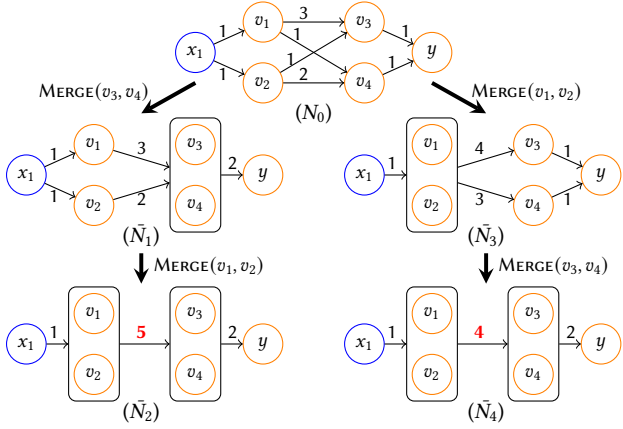

\begin{example}
\label{ex:abstract-order}
Let us consider the DNN $\nn_0$ shown in Figure~\ref{fig:abstract-order},
where all neurons are \inc.
Assume all biases are $0$.
By iteratively applying
the abstraction steps $\merge(v_3,v_4)$ and $\merge(v_1,v_2)$ (resp. $\merge(v_1,v_2)$ and $\merge(v_3,v_4)$) on
$\nn_0$,
we obtain the DNN  $\oa{\nn}_2$ (resp. $\oa{\nn}_4$) as depicted.
It is easy to see that $\oa{\nn}_2$ and $\oa{\nn}_4$ are different.
\end{example}

Example~\ref{ex:abstract-order} shows that different orders of
abstraction steps may result in different over-approximations. Conversely,
to reverse the effects of the abstraction
steps, the refinement steps must follow some specific order as well.

Recall that given a sequence $\tau$ of abstraction steps,
Lemma~\ref{lemma:permutation} enables it to be permutated into an
implicit sequence $\tau'$. When applied, $\tau'$ produces the same over-approximation as $\tau$.
Nevertheless,
Example~\ref{ex:abstract-order} indicates that the order of refinement steps should be constrained by the order of abstraction steps.
This motivates us to introduce a dependency relation between abstraction steps,
which restricts the order of refinement steps.

A straightforward dependency relation can be established on the
implicit sequence: each step depends on all steps applied before
it. Based on this dependency, refinement steps can be performed
backwards along the implicit sequence of abstraction steps, ensuring soundness by Lemmas~\ref{lemma:merge}
and~\ref{lemma:qfreeze}. We define the following dependency
relation which provides more flexibility for refinement.

\begin{definition}
\label{def:dependency}
Given an \emph{implicit} sequence of abstraction steps
$\tau=\beta_1\beta_2\cdots\beta_m$. The \emph{dependency relation} is
defined as follows:
\begin{enumerate}
  \item[(i)]\label{i:freeze} each step $\beta_i$ applied in layer
    $k_i$ depends on the \qfreeze{} step $\beta_j$ that happens in layer
    $k_j$, if $j<i$ and $k_j>k_i$;
  \item[(ii)]\label{i:merge} the \merge step $\beta_i$ that merges $v$
    and $v'$ in layer $k$, depends on (a) all \merge steps producing
    $v$ and $v'$ (if any), and (b) all \merge steps $\beta_j$ at both
    layers $k-1$ and $k+1$ if $j<i$.
\end{enumerate}
\end{definition}

Intuitively, Item~(i) ensures that all
\qfreeze{} steps at layer $k_j$ happen before any abstraction step at
layer $k_i < k_j$, so that Lemma~\ref{lemma:qfreeze} is
applicable. Item~(ii) guarantees well-definedness of the \merge
step on $v$ and $v'$, because it requires information provided by both
\merge steps producing $v$ and $v'$ as well as all \merge steps that
happened at adjacent layers.

A \emph{dependency graph} $\dg$ of $\tau$ is a directed acyclic graph derived
by the dependency relation, where the vertices are
abstraction steps and an edge from $\beta_i$ to $\beta_j$
exists if $\beta_i$ depends on $\beta_j$.  Thus, we get:
\begin{theorem}
\label{thm:refine}
Let $\oa{\nn}$ be the DNN constructed from $\nn$ by applying a sequence $\tau$
of abstraction steps, $\dg$ the dependency graph of the implicit order
$\tau'$ for $\tau$, and $\oa{\nn}'$ the DNN refined from
$\oa{\nn}$ by a refinement step $\gamma$. It holds that
$\oa{\nn}(\vec{x}) \ge \oa{\nn}'(\vec{x}) \ge \nn(\vec{x})$ for each
input $\vec{x}$, if $\gamma$ is the inverse of an abstraction step
$\beta$ in $\dg$ that has no incoming edges.
\end{theorem}

In summary, to refine an over-approximation $\oa{\nn}$, we first
construct the implicit sequence $\tau'$ and the corresponding
dependency graph $\dg$, %
respectively. We pick any abstraction
step in $\dg$ with no incoming edges, i.e. not depended by others, and
perform a corresponding refinement step. By Theorem~\ref{thm:refine},
the refined DNN $\oa{\nn}'$ in such a way ensures soundness, thus, partially solving challenge C3.

\begin{example}
Recall Example~\ref{ex:abstract-order} for the left part in
Figure~\ref{fig:abstract-order}. $\oa{\nn}_2$ is obtained from $\nn_0$
by sequence $\tau=[\merge(v_3,v_4), \merge(v_1,v_2)]$, whose implicit
sequence $\tau'=\tau$ by Lemma~\ref{lemma:permutation}. The
corresponding dependency graph $\dg$ contains only two vertices
$\merge(v_3,v_4)$ and $\merge(v_1,v_2)$ and an edge from
$\merge(v_1,v_2)$ to $\merge(v_3,v_4)$. To soundly refine
$\oa{\nn}_2$, only $\merge(v_1,v_2)$ can be picked to reverse.
\end{example}

\subsection{Counterexample-Guided Refinement}
\label{ss:guided-refinement}

Similar as the abstraction procedure, to address challenge C4,
the following two questions should be considered in the refinement procedure: \emph{which available
refinement step should be chosen}, and \emph{how many steps should be
performed}?

According to Theorem~\ref{thm:refine}, any abstraction step in the
dependency graph without incoming edges can be chosen for refinement.
Thus, there may exist several applicable refinement steps during an iteration of the refinement procedure.
Contrary to the abstraction strategy, we expect a refinement step to
restore more accuracy, which corresponds to the candidate abstraction
step that induced more inaccuracy.  Therefore, we estimate the gain of accuracy
for reversing a candidate abstraction step $s$ via the following profit function $\lossx{\cdot}$
w.r.t. the spurious counterexample $\ce$,
\[\small
\lossx{s} = \left\{
  \begin{array}{ll}
  \!\!\!|\sum_k{v^{(k)}(\ce)} - \oa{v}(\ce)|,
  & \!\!\!\!\!\mbox{if $s$ is \merge on $v^{(k)}$'s producing $\oa{v}$};\\
  \vspace{-3mm}\\
  \!\!\!|v(\ce) - \oa{b}(\oa{v})|,
  & \!\!\!\!\!\mbox{if $s$ is \qfreeze{} on $v$ producing $\oa{v}$;}
  \end{array}\right.
\]
where $v^{(k)}$'s denote the atomic neurons merged to $\oa{v}$,
$v(\ce)$ denotes the exact value of a neuron $v$ under the input $\ce$.

Based on $\lossx{\cdot}$, we propose a counterexample-guided refinement strategy (cf. Algorithm~\ref{alg:refinement}), which chooses a candidate
refinement step, by which we can regain the most accuracy.

\begin{algorithm}[t]
\caption{Counterexample-Guided Refinement Strategy}
\label{alg:refinement}
\begin{algorithmic}[1]
  \Require \!\!$\nn$, abstraction $\oa{\nn}$, dependency graph $\dg$,
  counterexample $\ce$
  \Ensure Refined network $\oa{\nn}'$, updated $\dg$
    \State Extract the set $C$ of applied abstraction steps from $\dg$
    \State $\maxerr \gets 0$, $\bestop\gets \bot$
    \For {each candidate step $s$ in $C$}
      \If {$\lossx{s} > \maxerr$}
        \State $\maxerr \gets \lossx{s}$,\; $\bestop\gets \mbox{refinement step for } s$
      \EndIf
    \EndFor
    \State Build $\oa{\nn}'$ from $\oa{\nn}$ by applying the refinement step $\bestop$
    \State Update dependency graph $\dg$
    \State \Return $\oa{\nn}'$, $\dg$
\end{algorithmic}
\end{algorithm}

However, the strategy in Algorithm~\ref{alg:refinement} does not guarantee that
the counterexample $\ce$ is ruled out by a single refinement
step, i.e., $\ce$ is still a counterexample of the refined
DNN $\oa{\nn}'$.  To solve
the second question for the refinement procedure, the routine in
Algorithm~\ref{alg:refinement} is repeated until $\ce$ is ruled out
from the refined DNN.

The counterexample-guided refinement strategy, coupled with the
terminating condition and refinement primitives, solves
challenges C3 and C4.

\section{Evaluation}\label{sec:experiment}
We implement our approach in a tool
\ourtool (\textbf{N}etwork \textbf{A}bstraction-\textbf{R}efinement for
\textbf{v}erification). \ourtool utilizes the promising symbolic
interval analysis tool \reluval~\cite{DBLP:conf/uss/WangPWYJ18} for
computing the bounds of neurons, while
the back-end verification engine can be configured with
any sound tool that can produce a counterexample when the verification problem does not hold.
To evaluate \ourtool, two promising tools
\marabou~\cite{DBLP:conf/cav/KatzHIJLLSTWZDK19} and
\planet~\cite{DBLP:conf/atva/Ehlers17} are integrated as the back-end
verification engines.  Both \marabou and \planet are sound and
complete, thus \ourtool is sound and complete.  \marabou is chosen as
the back-end verification engine, because it is the most efficient
tool among 13 tools participating in the 2nd International Verification
of Neural Networks Competition (\vnncomp)~\cite{vnn-comp21}
(cf. Table~5 in the summary and results of
\vnncomp~\cite{abs-2109-00498}).  \planet is
comparable to \marabou (even better on some
benchmarks)~\cite{DBLP:conf/cav/KatzHIJLLSTWZDK19}.

The experiments are designed to answer the following
research questions:
\begin{enumerate}
\item[RQ1:] \textbf{Effectiveness}. Can \ourtool boost the sound and
  complete verification tools \marabou and \planet?
\item[RQ2:] \textbf{Performance}. Does \ourtool outperform \cegarnn~\cite{DBLP:conf/cav/ElboherGK20},
the only work that supports structure-oriented CEGAR?
\end{enumerate}

We evaluate \ourtool on three widely-used benchmarks and datasets: the
DNNs from \acasxu~\cite{julian2019deep}, the DNNs trained by datasets
\mnist~\cite{lecun1998mnist} and
\cifar~\cite{krizhevsky2009learning}. The DNNs trained by \mnist and
\cifar are relatively large. Not all of complete methods are effective
to solve useful verification problems in acceptable time for
them~\cite{urban2021review}.

\textbf{\acasxu}, is a collision avoidance system built for unmanned
aircrafts. The system consists of 45 real-world DNNs, each of which
has 310 neurons including 5 inputs, 6 hidden layers and 5 outputs. The
inputs take normalized data from airborne sensors, representing the
relative position and speed of intruders. The outputs provide 5 kinds
of turning advisories to prevent the aircraft from collision.

\textbf{\mnist}, is a standard dataset for handwritten digit
recognition. The DNN used in our evaluation is provided by
\vnncomp. It contains 1306 neurons, including 2 hidden layers with 256
neurons per layer. Due to the $28\!\times\!28$ format of the images,
the input layer takes 784 pixels in greyscale and the output layer has
10 neurons producing the classification scores for the 10 possible
digits.

\textbf{\cifar}, is a colored image dataset that consists of 60000,
$32\!\times\!32$ RGB images in 10 classes (e.g., cat or dog).
The DNN used in our evaluation is collected from the
benchmarks of the ERAN toolset~\cite{eran-repo}. The size of its
hidden layers is $6\!\times\!100$.  The DNN has 3072 input neurons
representing the pixel values of the 3 color channels and 10 outputs
as the classification results. The total size is 3682.

The properties to be verified are the robustness of DNNs against
adversarial examples.
We verify whether the classification result for each input on a target
DNN remains the same after adding small perturbations onto that input,
where the perturbations are limited within a given threshold $\delta$
using the $L_\infty$ norm~\cite{Carlini017}.

All experiments are run on a Linux server with two Intel Xeon
Sliver-4214 CPUs and 64 GB memory. The timeout is set to 1 hour for
each \acasxu verification problem as in \vnncomp, whereas 10 hours for
each verification problem on \mnist and \cifar DNNs.

The experimental results show that our approach is able to significantly boost
the performance and scalability of both \marabou and \planet
and significantly outperforms \cegarnn, the only work that supports structure-oriented CEGAR for DNN verification.

\subsection{RQ1: Effectiveness Evaluation}

\noindent{\bf Setup.}
To answer {\bf RQ1},
we evaluate the effectiveness of \ourtool on the relatively large
networks trained by \mnist and \cifar.  Two \ourtool configurations
\ourtoolm and \ourtoolp are set up respectively with \marabou and
\planet as their back-end verification engines.
\ourtoolm and \ourtoolp are then compared with \marabou and \planet,
respectively, on both the \mnist and the \cifar networks.
For each network, we verify robustness against 4 perturbation thresholds $\delta$ ranging: from $0.02$ to $0.05$ for the
\mnist network, and from $0.001$ to $0.004$ for the \cifar
network. These thresholds are selected since the robustness threshold
is close to $0.06$ for the \mnist network, and $0.005$ for
\cifar network, thus yielding interesting yet hard robustness verification problems. There are 25 and 12 verification problems for each
perturbation threshold $\delta$ respectively for \mnist and \cifar.

\smallskip
\noindent{\bf Results on the \mnist network.}  Table~\ref{tab:mnist}
(upper part) reports the results on the
\mnist network, where the average verification time in seconds ("Avg. time") and the
numbers of successfully verified problems ("\#Verified") by each tool
are depicted.

\begin{table}\small
\caption{Effectiveness Evaluation of our Approach on \marabou and \planet}\vspace{-1mm}
  \centering\setlength\tabcolsep{10pt}
\begin{tabular}{|c|r|r|r|r|r|}\cline{3-6}
 \multicolumn{2}{c|}{}
  & \multicolumn{1}{c|}{Avg. time (s)}
  & \multicolumn{1}{c|}{\#Verified}
  & \multicolumn{1}{c|}{Avg. time (s)}
  & \multicolumn{1}{c|}{\#Verified} \\
  \hline
  Dataset  &\multicolumn{1}{c|}{$\delta$} 
  & \multicolumn{2}{c|}{\ourtoolm}
  & \multicolumn{2}{c|}{\marabou} 
    \\ \hline 
  \multirow{4}*{{\mnist}}
  & 0.02
  & 99.54 \hfill (\tinc{65.0\%})   & 25 \hfill ( - )
  & 60.34   & 25 
  \\ %
  & 0.03
  & 971.70 \hfill (\tdec{86.3\%}) & 25 \hfill (\ninc{12\%})
  & 7084.99   & 22 
  \\ %
  & 0.04
  & 2458.05 \hfill (\tdec{85.6\%}) & 25 \hfill (\ninc{40\%})
  & 17063.42   & 15 
  \\ %
  & 0.05
  & 8056.73 \hfill (\tdec{71.7\%}) & 20 \hfill (\ninc{52\%})
  & 28465.46   & 7 
  \\ \hline%
  \multirow{4}*{{\cifar}}
    & 0.001
    & 1581.66 \hfill (\tdec{82.1\%}) & 12 \hfill ( - )
    & 8824.25   & 12 
    \\ %
    & 0.002
    & 12545.33 \hfill (\tdec{56.0\%}) & 8 \hfill (\ninc{33\%})
    & 28515.69   & 4 
    \\ %
    & 0.003
    & 15241.70 \hfill (\tdec{57.7\%}) & 7 \hfill (\ninc{58\%})
    & 36000.00   & 0 
    \\ %
    & 0.004
    & 15200.91 \hfill (\tdec{57.8\%}) & 7 \hfill (\ninc{58\%})
    & 36000.00   & 0 
    \\  \hline 
  \hline
  Dataset  &\multicolumn{1}{c|}{$\delta$}
  & \multicolumn{2}{c|}{\ourtoolp}
  & \multicolumn{2}{c|}{\planet}
    \\ \hline 
        \multirow{4}*{{\mnist}}
  & 0.02
  & 118.91 \hfill (\tdec{67.5\%}) & 25 \hfill ( - )
  & 366.39   & 25
  \\ %
  & 0.03
  & 259.54 \hfill (\tdec{78.0\%}) & 25 \hfill ( - )
  & 1182.32   & 25
  \\ %
  & 0.04
  & 454.14 \hfill (\tdec{60.3\%}) & 25 \hfill ( - )
  & 1143.16   & 25
  \\ %
  & 0.05
  & 585.59 \hfill (\tdec{54.1\%}) & 25 \hfill ( - )
  & 1275.57   & 25
  \\ \hline%
  \multirow{4}*{{\cifar}}
    & 0.001
    & 12682.02 \hfill (\tdec{63.4\%}) & 8 \hfill (\ninc{58\%})
    & 34641.92    & 1
    \\ %
    & 0.002
    & 19097.04 \hfill (\tdec{47.0\%}) & 7 \hfill (\ninc{58\%})
    & 36000.00    & 0
    \\ %
    & 0.003
    & 15249.89 \hfill (\tdec{57.6\%}) & 3 \hfill (\ninc{25\%})
    & 36000.00   & 0
    \\ %
    & 0.004
    & 21187.41 \hfill (\tdec{41.1\%}) & 3 \hfill (\ninc{25\%})
    & 36000.00   & 0 \\ \hline
\end{tabular}
\label{tab:mnist}
\end{table}

We can observe that \ourtoolm and \ourtoolp can solve more problems and
in general are much faster than  \marabou and \planet, in particular on hard verification problems.
This indicates that our approach is able to significantly boost the performance of both \marabou and \planet
in most cases on the \mnist network. The best
improvements appear when $\delta=0.03$, reducing $86.3\%$ and $78.0\%$
verification time respectively for \marabou and \planet.

In detail, \marabou appears to have a good
performance when the perturbation threshold $\delta$ is small, as the
solving space of the verification problem is small.  It begins to time
out (10 hours) when $\delta\geq 0.03$, and the numbers of solved problems become
less and less with the increase of $\delta$. When $\delta=0.05$, it
only successfully solves 7 ($28\%$) problems out of all the 25 verification problems.
In contrast, \ourtoolm is able to solve most of the problems
for all perturbation thresholds except for $\delta=0.05$. When
$\delta=0.05$, it solves 20 problems, namely $52\%$ ($=(20-7)/25$)
more than \marabou. In addition, \ourtoolm spends $71.7\%$ ($\approx
(28465.46-8056.73)/28465.46$) less verification time on average when
$\delta=0.05$.

Compared over \planet which is able to solve all the \mnist
verification problems as depicted in Table~\ref{tab:mnist}, \ourtoolp
also solves all the problems, meanwhile spends less time on average
for all the perturbation thresholds, reducing at least 54.1\% verification time.

Finally, we should emphasize that verified robustness with large
perturbation threshold $\delta$ is more interesting in practice.

\begin{figure}[t]
  \!\!\!\!%
  \subfigure[\ourtoolm vs. \marabou]{
    \includegraphics[width=0.35\columnwidth]
                    {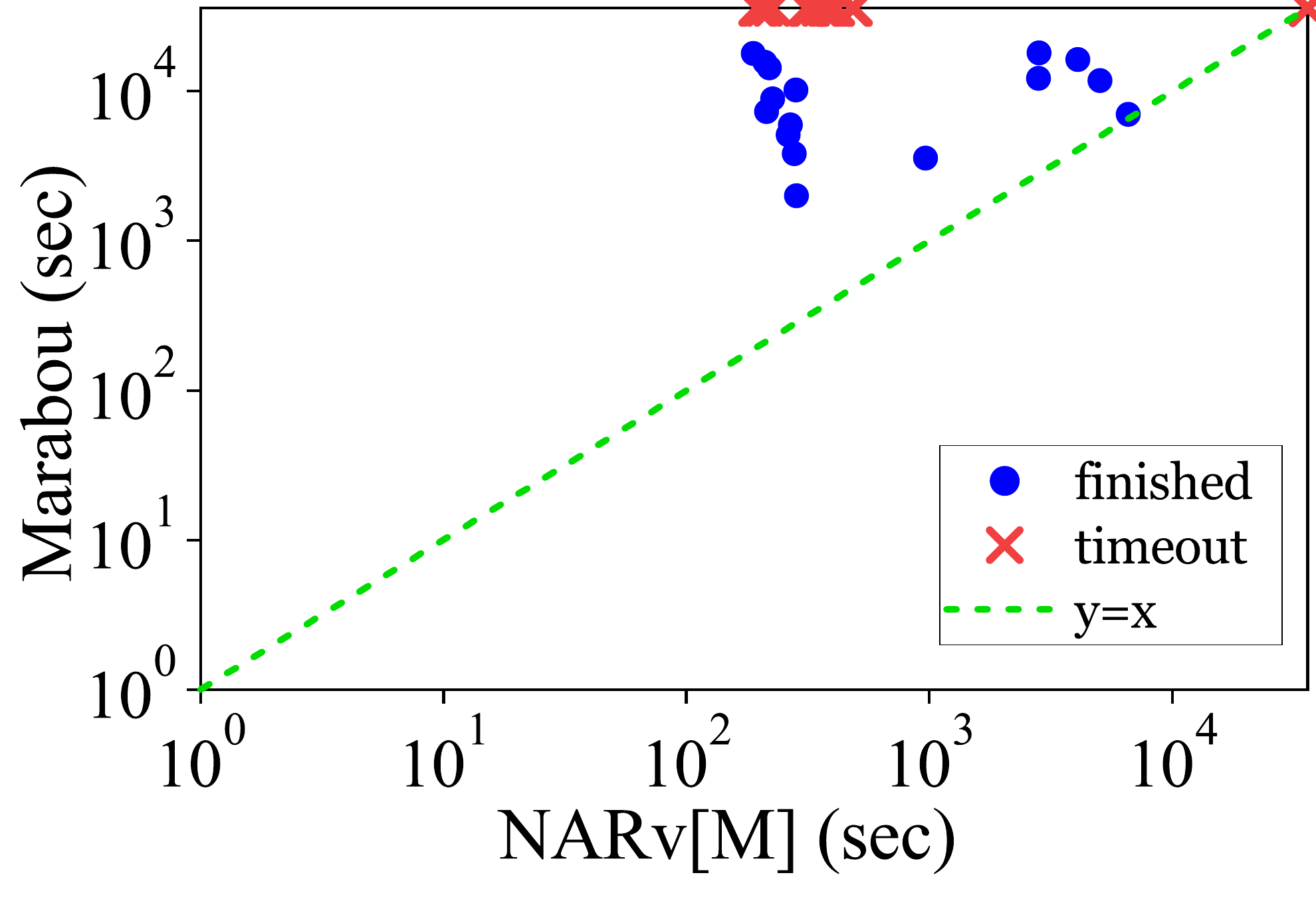}
    \label{subfig:cifar600-marabou}
  }%
  \qquad
  \subfigure[\ourtoolp vs. \planet]{
    \includegraphics[width=0.35\columnwidth]
                    {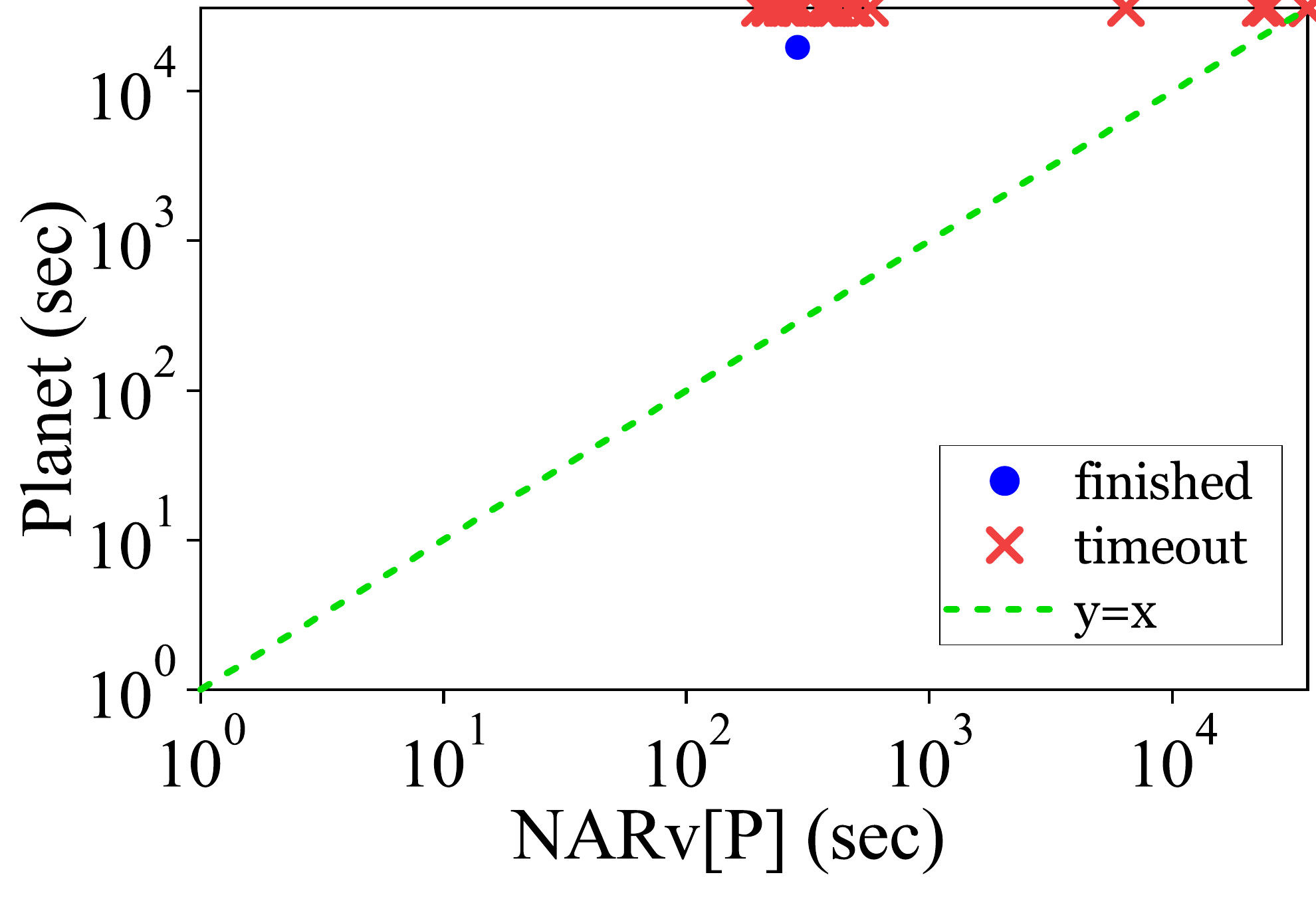}
    \label{subfig:cifar600-planet}
  }%
  \vspace{-2mm}
  \caption{Verification Time on \hide{$4\times100$} \cifar Network
    (logscale), with \marabou, \planet and their Boosted Versions}
  \label{fig:cifar600} %
\end{figure}

\smallskip
\noindent{\bf Results on the \cifar network.}  The experimental
results on the \cifar network are depicted in Figure~\ref{fig:cifar600},
where Figure~\ref{subfig:cifar600-marabou} (resp. Figure~\ref{subfig:cifar600-planet})
compares the verification time required by \ourtoolm and \marabou (resp. \ourtoolp
and \planet) for each verification
problem, and the x-axis and  y-axis refer to the  time (in
seconds, logscale) spent by \ourtoolm and \marabou (resp. \ourtoolp and \planet).
The blue dots represent the solved problems and the red crosses on the top and right borders are the cases
where \marabou (resp. \planet) and \ourtoolm (resp. \ourtoolp)  run out of time.
Therefore, the dots and crosses
distributed above the green line indicate the cases where \ourtoolm (resp. \ourtoolp) is
faster.

It is easy to observe that most of the dots and
crosses are situated strictly above the green line in both
Figures~\ref{subfig:cifar600-marabou}
and~\ref{subfig:cifar600-planet}, indicating that our approach can
significantly boost \marabou and \planet in most cases. We should
point out that \planet can hardly solve any verification problems in
the \cifar network.  Nevertheless, boosted by our approach,
\ourtoolp is still able to solve many verification problems within the
time limit (10 hours).

Table~\ref{tab:mnist} (lower part) gives the detailed results.  It
shows that \planet fails to solve any problems before timing out when
$\delta \geq 0.002$. It successfully solves only 1 ($2.1\%$) problem
out of all 48 problems for 4 different perturbation
thresholds. Boosted by our approach, \ourtoolp is able to
solve 7 ($58\%$) and 3 ($25\%$) problems respectively when
$\delta=0.002$ and $\delta\geq 0.003$, succeeding in $41.7\%$
($\approx (8+7+3+3-1)/48$) more in total.  On the other hand,
\ourtoolm is able to solve $37.5\%$ more problems in total than
\marabou.

\begin{tcolorbox}[size=title]
{\textbf{Answer to RQ1:}}
\ourtool can significantly boost
the performance of two promising tools \marabou and \planet.
\end{tcolorbox}

\subsection{RQ2: Performance Evaluation}

\noindent{\bf Setup.}
To answer {\bf RQ2}, we compare \ourtool over \cegarnn,
both of which use the same back-end verification engine \marabou.
\cegarnn provides two abstraction strategies, named
\emph{indicator-guided abstraction} and \emph{abstraction to
saturation}, where the former iteratively merges two neurons until some pre-sampled input violates the given property
and the latter aggressively and iteratively merges two neurons producing the smallest over-approximation.
We refer to \cegarnn with those abstraction strategies as \cegarnni and \cegarnns,
respectively. \ourtool is then compared with both of them using the same
benchmark \acasxu as adopted in~\cite{DBLP:conf/cav/ElboherGK20}.
The perturbation threshold $\delta$
is set from
$0.01$ to $0.04$.

\begin{figure}[!t]
  \!\!\!\!%
  \subfigure[$\delta = 0.01$]{
    \includegraphics[width=0.35\columnwidth]
                    {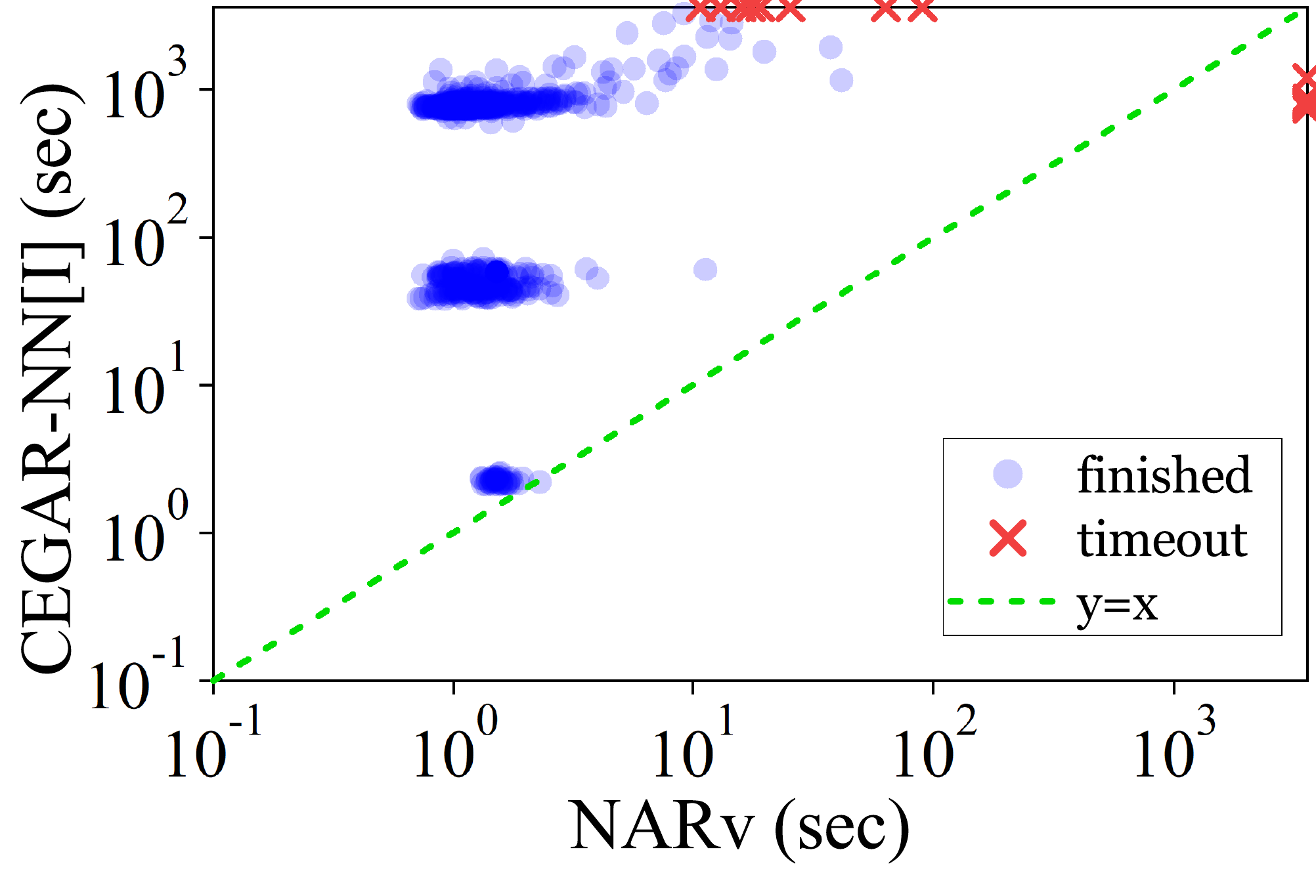}
    \label{subfig:cegarnni-001}
  }%
  \qquad
  \subfigure[$\delta = 0.02$]{
    \includegraphics[width=0.35\columnwidth]
                    {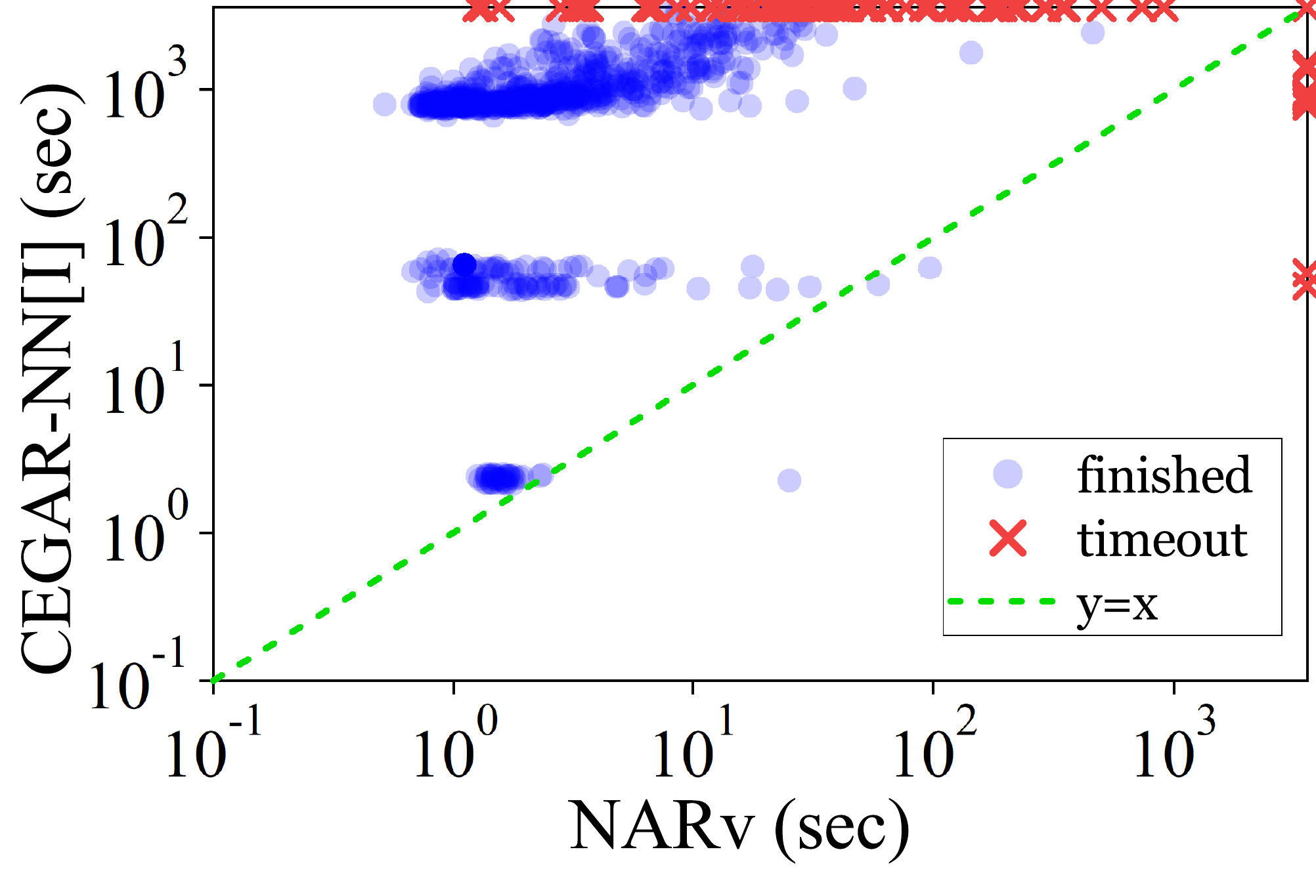}
    \label{subfig:cegarnni-002}
  }\\
  \!\!\!\!%
  \subfigure[$\delta = 0.03$]{
    \includegraphics[width=0.35\columnwidth]
                    {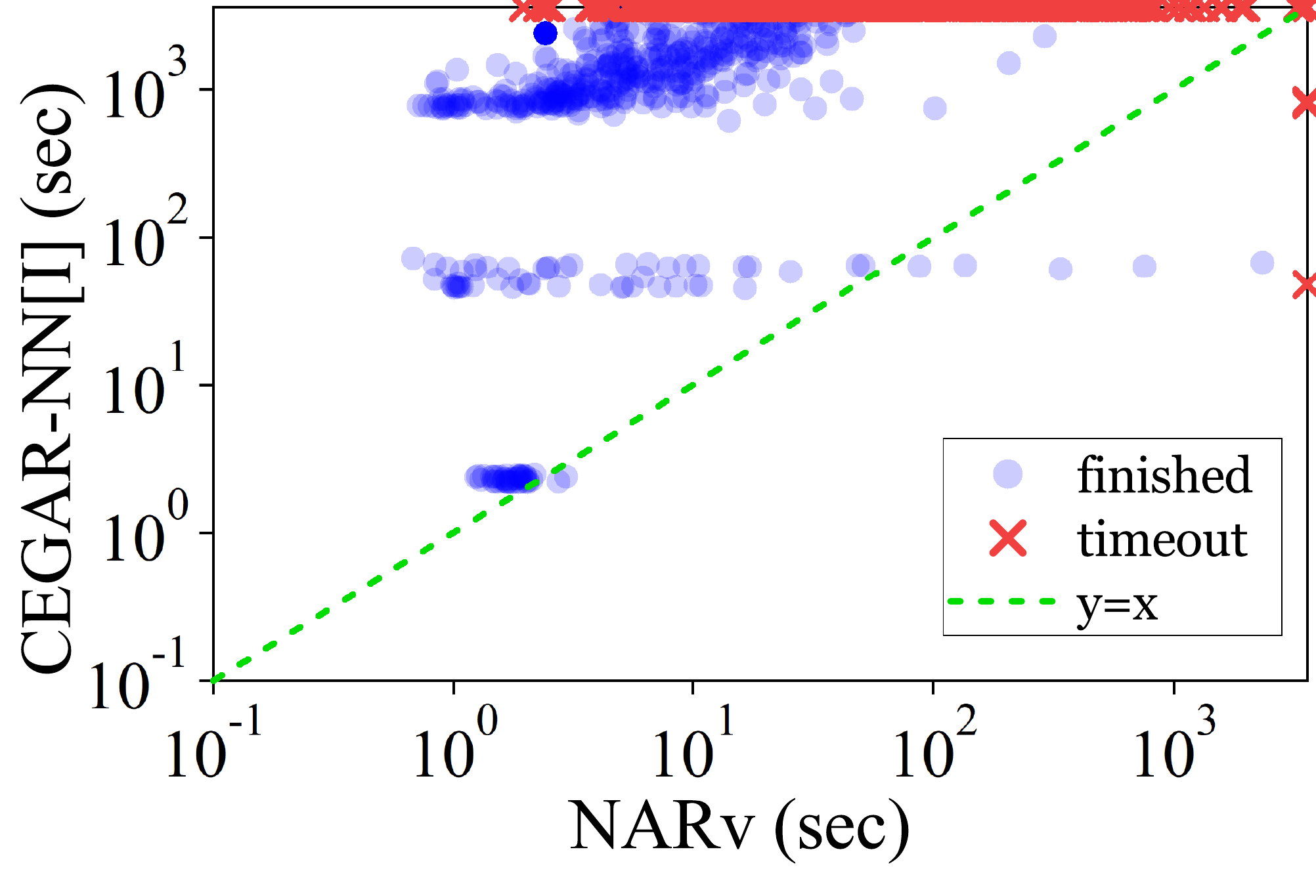}
    \label{subfig:cegarnni-003}
  }%
  \qquad\subfigure[$\delta = 0.04$]{
    \includegraphics[width=0.35\columnwidth]
                    {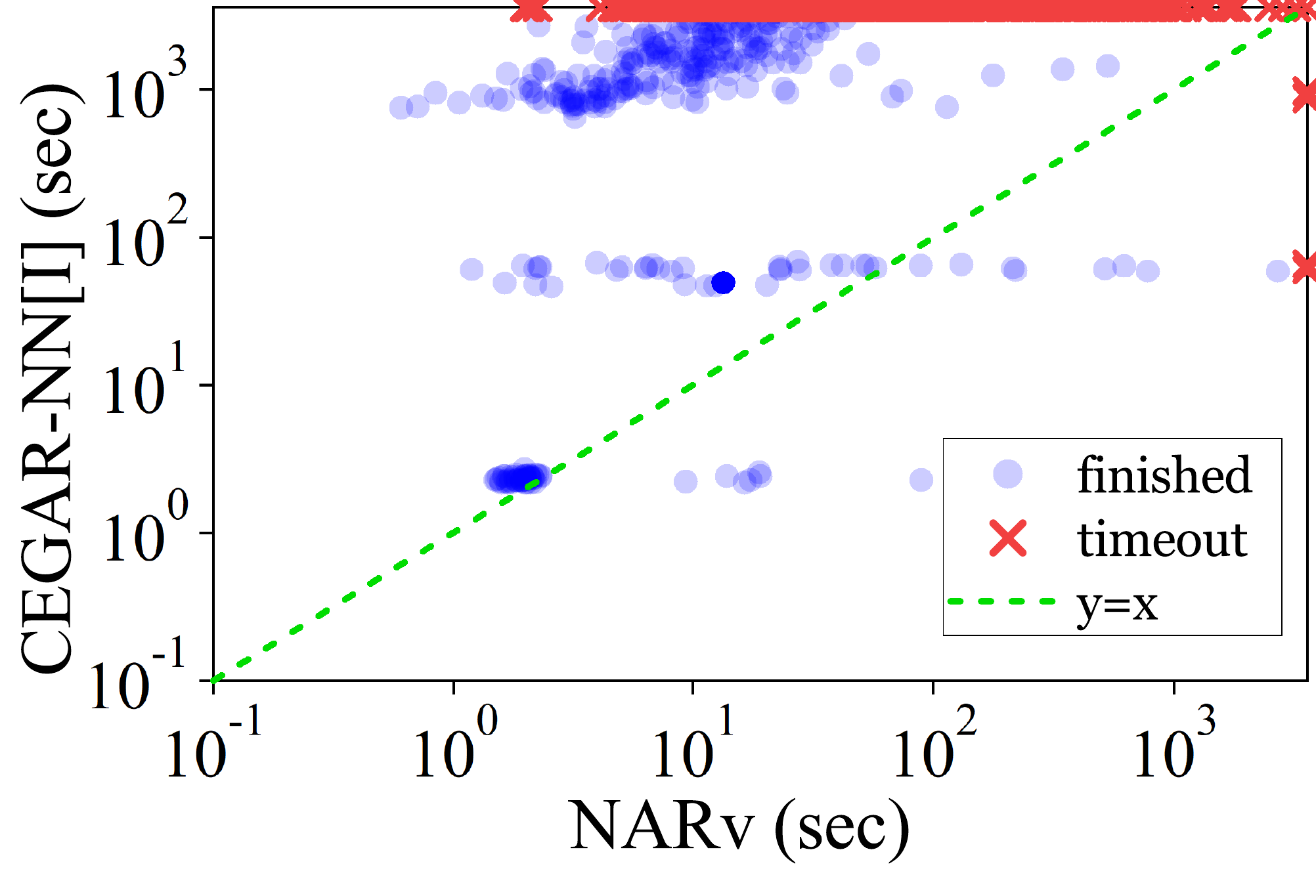}
    \label{subfig:cegarnni-004}
  }
  \vspace{-2mm}\caption{Comparison to \cegarnni on \acasxu (logscale)}
\label{fig:cegarnni}  %
\end{figure}

\begin{figure}[!t]
  \!\!\!\!%
  \subfigure[$\delta = 0.01$]{
    \includegraphics[width=0.35\columnwidth]
                    {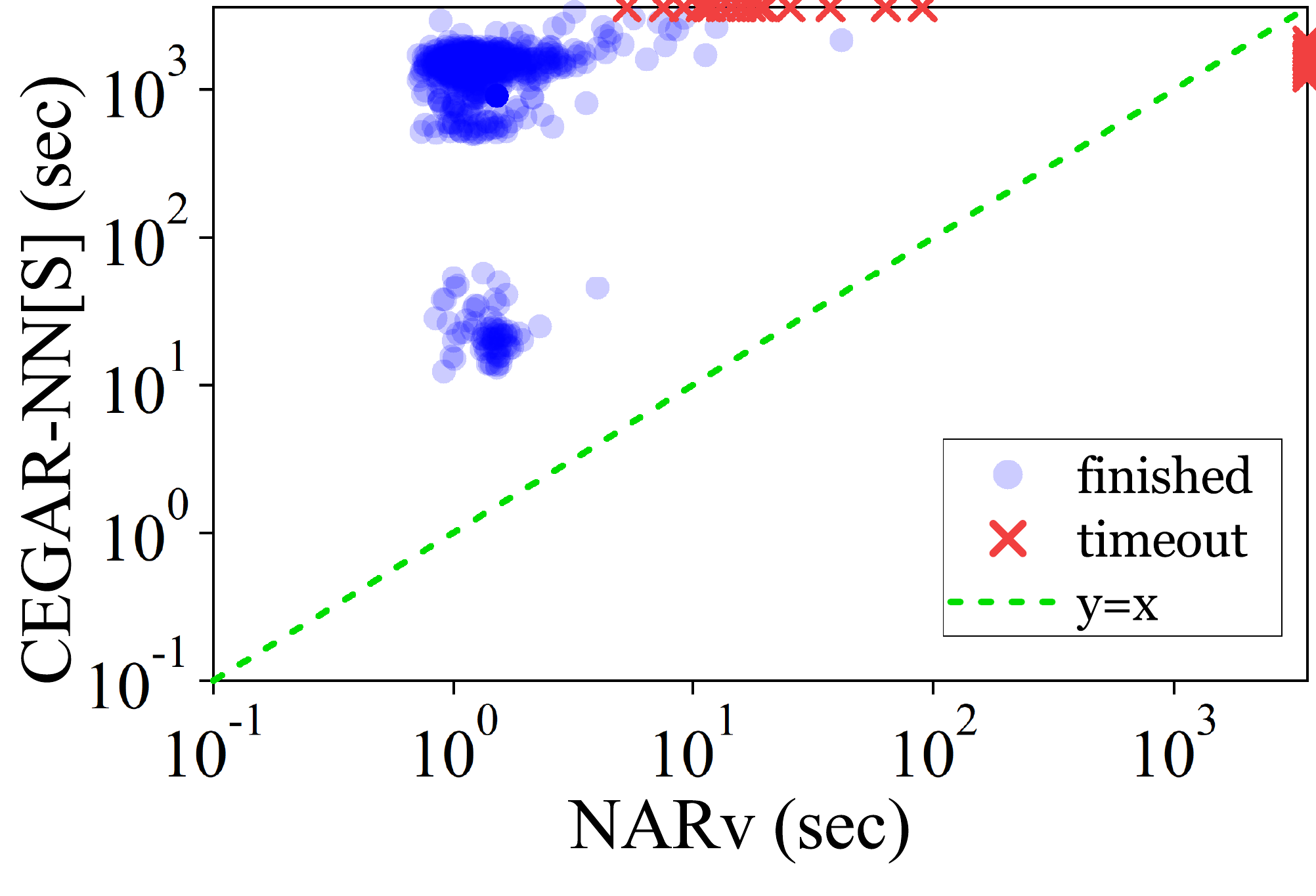}
    \label{subfig:cegarnni-001}
  }%
  \qquad
  \subfigure[$\delta = 0.02$]{
    \includegraphics[width=0.35\columnwidth]
                    {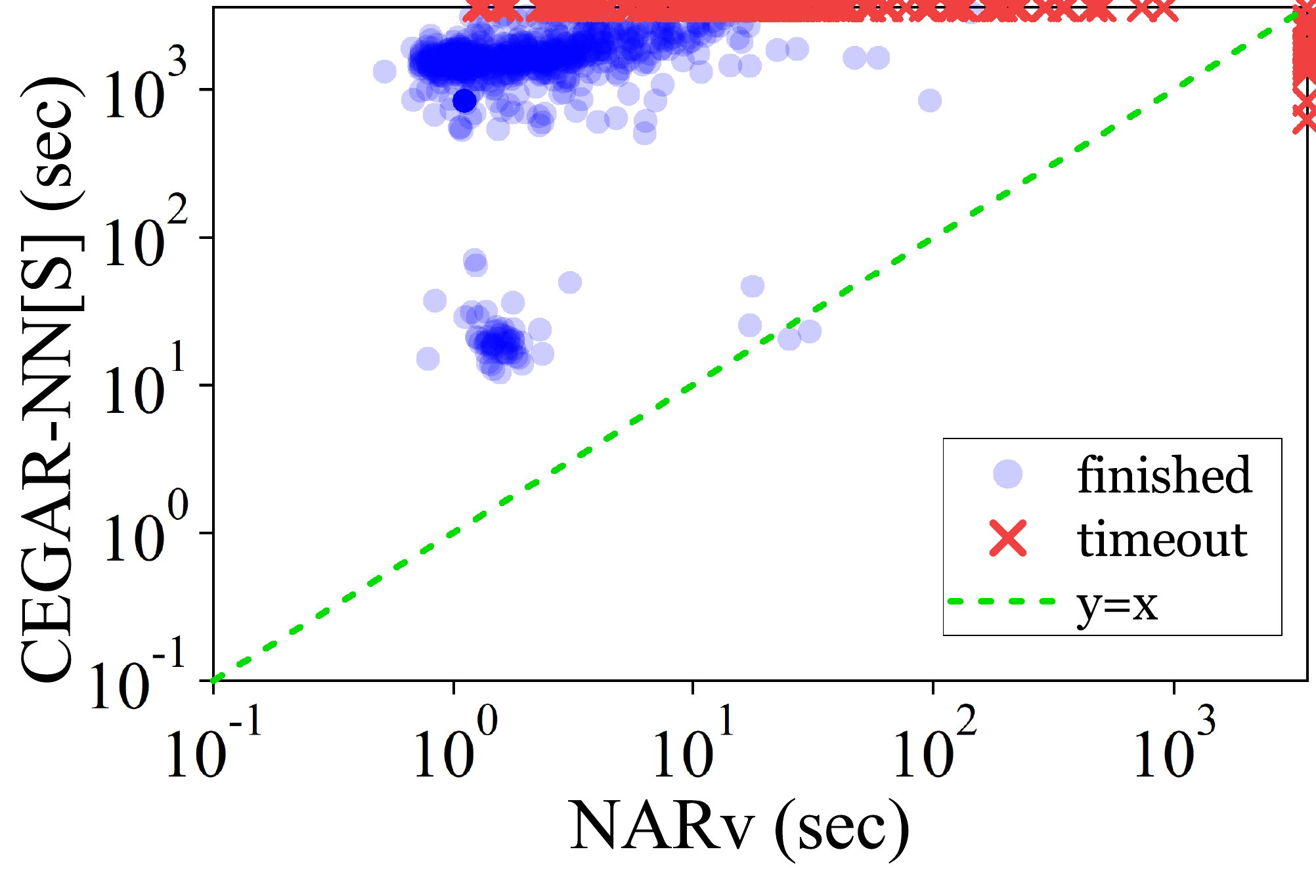}
    \label{subfig:cegarnni-002}
  }\\
  \!\!\!\!%
  \subfigure[$\delta = 0.03$]{
    \includegraphics[width=0.35\columnwidth]
                    {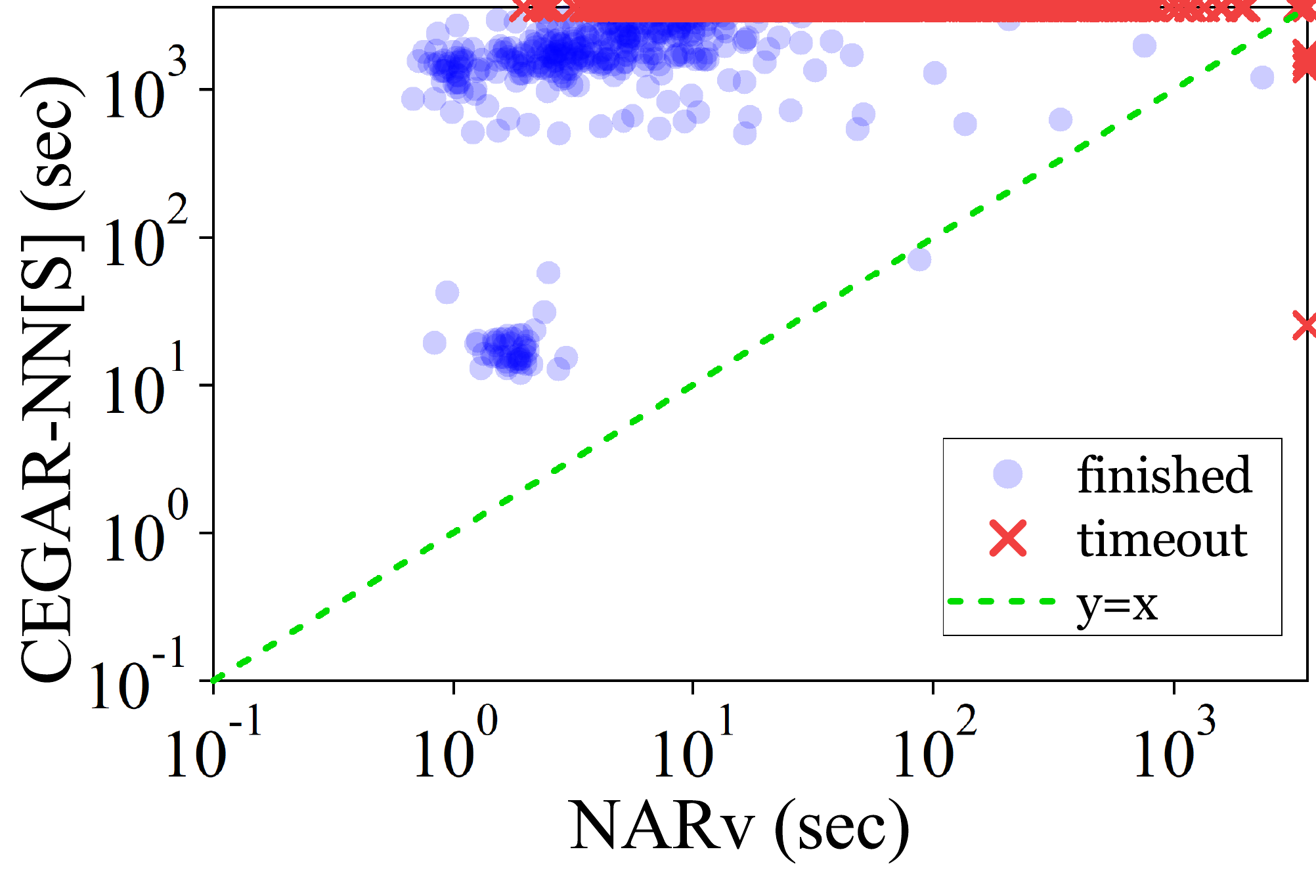}
    \label{subfig:cegarnni-003}
  }%
  \qquad\subfigure[$\delta = 0.04$]{
    \includegraphics[width=0.35\columnwidth]
                    {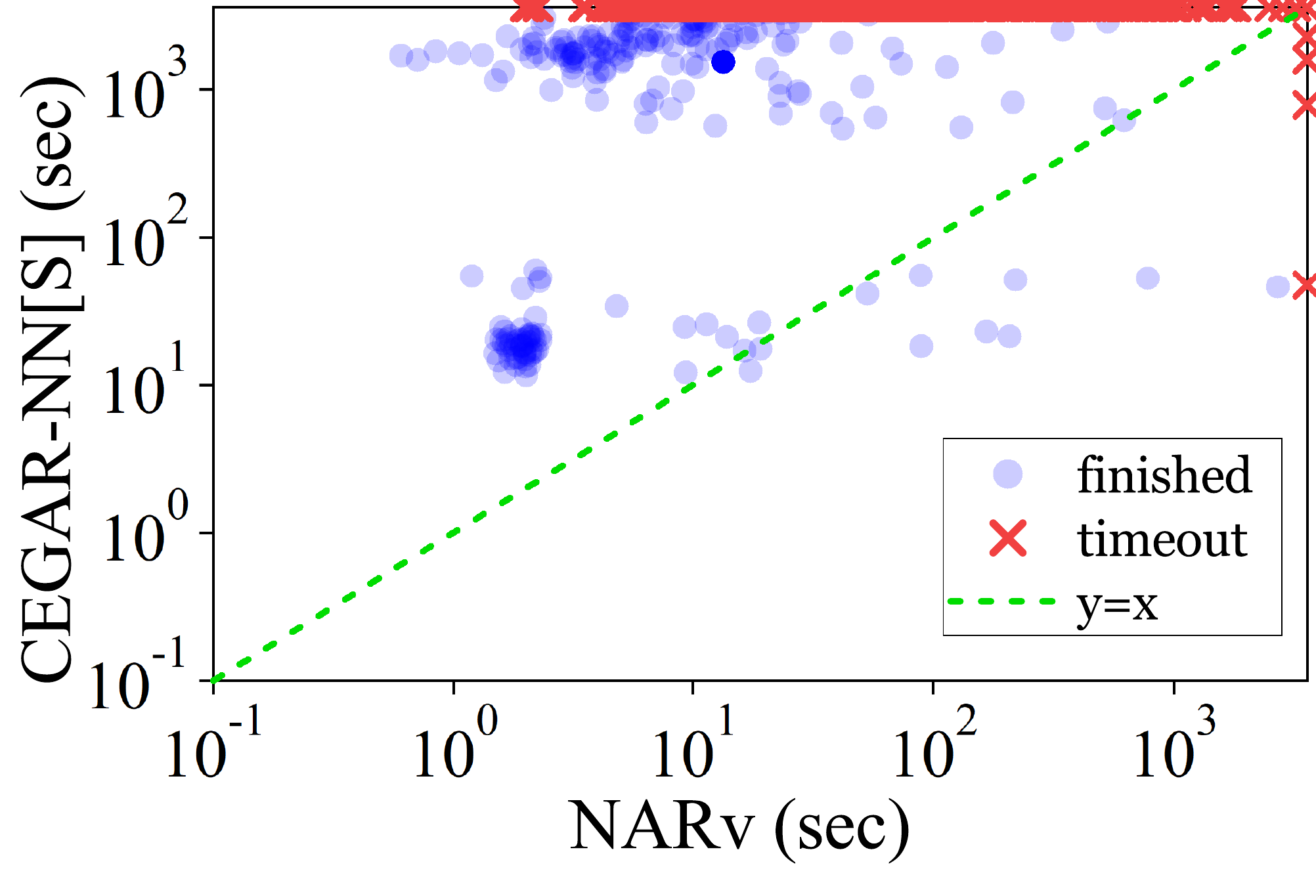}
    \label{subfig:cegarnni-004}
  }
  \vspace{-2mm}\caption{Comparison to \cegarnns on \acasxu (logscale)}
\label{fig:cegarnns} 
\end{figure}

\smallskip
\noindent{\bf Results.}
Figure~\ref{fig:cegarnni} depicts the comparison of the
verification time between \ourtool and \cegarnni for
each  problem.  We observe that most of the dots and
crosses are located above the green line for all perturbation
thresholds, indicating that \ourtool is significantly more
efficient than \cegarnni on most verification problems.
The comparison of the verification time between \ourtool and \cegarnns
is illustrated in Figure~\ref{fig:cegarnns}, where a similar
conclusion can be drawn.

Table~\ref{tab:cegarnn} reports the details of all the results, where
columns ("Avg. time") and ("\#Verified") are the same as above,
and columns ("Avg. size") give the average sizes of hidden
neurons in the abstract networks when the tools successfully verify
the problems.

\begin{table}[!htb]\small
  \caption{Abstraction and Verification Results for \acasxu
    Networks}\vspace{-2mm}  \centering\setlength\tabcolsep{10pt}
  \label{tab:cegarnn}
\begin{tabular}{|c|c|c|c|} \cline{2-4}%
 \multicolumn{1}{c|}{} & \multicolumn{1}{c|}{\ourtool}
  & \multicolumn{1}{c|}{\cegarnni}
  & \multicolumn{1}{c|}{\cegarnns} 
    \\ \hline
   \multicolumn{1}{|c|}{$\delta$} & \multicolumn{3}{c|}{Avg. time (s)} 
  \\ \hline
  0.01
  & \textbf{49.82} %
  & 629.16 \hfill (12.6\x)     & 1373.35 \hfill ~(27.6\x)
  \\  %
  0.02
  & \textbf{88.10} %
  & 1184.92 \hfill ~(13.4\x)    & 1980.36 \hfill (22.5\x)
  \\  %
  0.03
  & \textbf{105.04} %
  & 2331.45 \hfill (22.2\x) & 2835.72 \hfill (27.0\x)
  \\  %
  0.04
  & \textbf{169.46} %
  & 2742.36 \hfill (16.2\x) & 3034.63 \hfill (17.9\x) \\ \hline \hline
   \multicolumn{1}{|c|}{$\delta$}  & \multicolumn{3}{c|}{\#Verified}
  \\ \hline
  0.01
 & 888 \hfill (98.7\%) & \textbf{892} \hfill (99.1\%) & 884 \hfill (98.2\%) 
  \\  %
  0.02
  & \textbf{881} \hfill (97.9\%) & 815 \hfill (90.6\%) & 731 \hfill (81.2\%) 
  \\  %
  0.03
  & \textbf{893} \hfill (99.2\%) & 497 \hfill (55.2\%) & 368 \hfill (40.9\%) 
  \\  %
  0.04
  & \textbf{886} \hfill (98.4\%) & 327 \hfill (36.3\%) & 233 \hfill (25.9\%) \\ \hline \hline
   \multicolumn{1}{|c|}{$\delta$}  & \multicolumn{3}{c|}{Avg. size} \\ \hline
    0.01
  & \textbf{271} \hfill (\tdec{10\%})
  & 860 \hfill (\tinc{187\%})
  & 828 \hfill (\tinc{176\%})
  \\  %
  0.02
  & \textbf{210} \hfill (\tdec{30\%})
  & 882 \hfill (\tinc{194\%})
  & 824 \hfill (\tinc{175\%})
  \\  %
  0.03
   & \textbf{212} \hfill (\tdec{29\%})
  & 887 \hfill (\tinc{196\%})
  & 784 \hfill (\tinc{161\%})
  \\  %
  0.04
    & \textbf{229} \hfill (\tdec{24\%})
  & 884 \hfill (\tinc{195\%})
  & 684 \hfill (\tinc{128\%})
  \\ \hline
\end{tabular}
\end{table}

Table~\ref{tab:cegarnn} shows that \ourtool is much faster than both
\cegarnni and \cegarnns.  Particularly,
when $\delta=0.01$, \ourtool is on average $11.6$ ($\approx
(629.16-49.82)/49.82$) times faster than \cegarnni, and $26.6$
($\approx (1373.35-49.82)/49.82$) times faster than \cegarnns.
For the network sizes, recall that there are 300 hidden neurons in
each given \acasxu network. \ourtool succeeds in reducing the sizes of
hidden neurons for all perturbation thresholds, the best reduction
being $30\%$ ($=(300-210)/300$) on average when $\delta=0.02$.
However, neither of the two abstraction strategies of \cegarnn has progress in
size reduction.  One of the main reasons is that \cegarnn preprocesses
the target DNN by first quadrupling the size, leading to a heavy load
for the abstraction procedure. As a structure-oriented abstraction approach, the
abstraction of \cegarnn appears less practical.  A well-designed
abstraction procedure leads to a better verification performance. As a
result, \ourtool is able to solve the most problems in a reasonable
time among the three tools when $\delta\geq 0.02$, and is comparable
to the other two when $\delta = 0.01$.  In particular, \ourtool
successfully verifies $72.6\%$ ($\approx (886-233)/900$) more problems
than \cegarnns when $\delta=0.04$, spending only $5.6\%$ ($\approx
169.46/3034.63$) of time on average.

\begin{tcolorbox}[size=title]
{\textbf{Answer to RQ2:}} \ourtool significantly outperforms {\cegarnn}, the
only work that supports structure-oriented CEGAR.
\end{tcolorbox}

\subsection{Threats to Validity}
Our approach is designed for fully connected feedforward deep neural networks with the \relu activation function.
It could cope with any monotonic activation
functions such as sigmoid and tanh~\cite{PW22}, but we have
not yet evaluated the effectiveness.
There are deep neural networks such as
convolutional and recurrent ones.
The former could be verified by equivalently transforming into fully connected ones~\cite{abs-1712-01252},
but utilizing the particular properties of
convolutional constructs to build abstraction is certainly an
interesting future work. The feasibility has been shown
in the concurrent work~\cite{DBLP:journals/corr/abs-2201-01978}.
However, it remains an open problem for recurrent neural networks.

The verification engines \marabou and \planet adopted in the experiments
are primarily based on SMT solving and thus complete but computational expensive.
Our structure-oriented CEGAR approach is proposed to boost them.
We have not evaluated with the abstract interpretation approaches,
which can be seen as computation-oriented abstraction, thus are incomplete, e.g., ~\cite{DBLP:conf/sp/GehrMDTCV18,DBLP:journals/pacmpl/SinghGPV19}.
Although refinement techniques also have been proposed, e.g., input refinement~\cite{DBLP:conf/uss/WangPWYJ18},
such computation-oriented abstraction-refinement frameworks are orthogonal to
our structure-oriented one.
Investigating the synergy between them is left for future work,
for instance, using them at a back-end verification engine in our approach.
To reduce this threat, we use the most efficient
one in {\vnncomp} (i.e., \marabou) and another comparable tool \planet.

Our approach may fail to boost DNN verification when the verification problems
are very easy. More effective heuristics should be
added to improve the efficiency for such verification problems.
For instance, abstract interpretation based approaches could be leveraged
before applying our approach. Nevertheless,
our approach can significantly boost the verification of hard problems
which are arguably more interesting and important in practice.

\section{Related Work}\label{sec:related}

A large and growing body of work studies heuristic search or other dynamic analysis techniques
to test robustness of neural networks, e.g.,~\cite{Carlini017,MaLLZG18,PeiCYJ19,SunWRHKK18,TianPJR18,MaJZSXLCSLLZW18},
cf.~\cite{ZhangHML22} for a survey.
They are often effective in finding adversarial examples
or violations of properties, as opposed to proving the absence of violations.
Thus, they are orthogonal to formal verification considered in this work.

The simple and earlier neural network verification makes use of constraint solving,
where a verification problem is reduced to the solving of constraints, e.g.,
SAT/SMT solving~\cite{DBLP:conf/cav/HuangKWW17,DBLP:conf/cav/PulinaT10, DBLP:conf/atva/Ehlers17,
DBLP:conf/cav/KatzBDJK17,DBLP:conf/cav/KatzHIJLLSTWZDK19},
LP/MILP solving~\cite{DBLP:conf/icml/WongK18,DBLP:conf/cvpr/LinYCZLLH19,DBLP:journals/corr/LomuscioM17,DBLP:journals/corr/abs-1711-07356,DBLP:conf/nfm/DuttaJST18}.
Most of these works focus on DNNs with
the ReLU activation function. Although these techniques are often sound and complete, %
they are limited in scalability.  Our work targets this scalability
problem and boosts existing DNN verification techniques by reducing
network sizes.

To improve the scalability, some other DNN verification techniques
utilize the idea of abstraction through abstract
interpretation~\cite{DBLP:conf/popl/CousotC77}. These techniques
include \aisqr~\cite{DBLP:conf/sp/GehrMDTCV18}, \deepz~\cite{DBLP:conf/nips/SinghGMPV18}, \deeppoly~\cite{DBLP:journals/pacmpl/SinghGPV19}, \reluval~\cite{DBLP:conf/uss/WangPWYJ18}, \nnv~\cite{DBLP:conf/fm/TranLMYNXJ19}, \deepsrgr~\cite{DBLP:conf/tacas/YangLLHWSXZ21}
and so on. The key idea is to use well-designed numerical abstract
domains, such as boxes~\cite{DBLP:conf/popl/CousotC77},
zonotopes~\cite{DBLP:conf/cav/GhorbalGP09} and
polyhedra~\cite{DBLP:conf/popl/CousotH78}, to over-approximate the
computations on sets of inputs in the target DNNs.  Due to the
over-approximation, these techniques are
incomplete.
To improve accuracy, some of them incorporate refinement
strategies, such as input region
splitting~\cite{DBLP:conf/uss/WangPWYJ18} and output bound
tightening~\cite{DBLP:journals/pacmpl/SinghGPV19,DBLP:conf/tacas/YangLLHWSXZ21}.
Different from them, the CEGAR-based approach adopted in
this work is based on the structure of DNNs, rather than on the computations,
thus is orthogonal to them.

There are other structure-oriented techniques. %
The abstraction technique proposed in~\cite{DBLP:conf/nips/PrabhakarA19}
represents the network weights via intervals and merges
randomly chosen neurons by merging intervals.
A k-means clustering algorithm is leveraged to
cluster neurons~\cite{DBLP:conf/atva/AshokHKM20}.
In contrast to ours, they rely on specific underlying
verification engines and do not provide refinement mechanisms
when some spurious counterexamples are reported, thus are incomplete.
Gokulanathan et al.~\cite{DBLP:conf/nfm/GokulanathanFMB20} proposed to identify and remove inactive neurons 
via DNN verification solving without harming accuracy.
But it has to invoke a large number of DNN verification queries.   

The closest work to ours is CEGAR-NN~\cite{DBLP:conf/cav/ElboherGK20}
which proposed the first structure-oriented CEGAR approach
for DNN verification. Inspired by CEGAR-NN, although the high-level CEGAR framework
is the same, our work makes three significant technical contributions:
(i) our preprocessing at most \emph{doubles} the network size, whilst \cegarnn at most \emph{quadruples} the
 size, leaving a heavy load to the following abstraction
procedure as shown in our experimental results; (ii) we provide two complementary
abstraction (resp. refinement) primitives that can induce less
inaccuracy (resp. regain more accuracy) whereas \cegarnn only provides one abstraction (resp. refinement) primitive;
 (iii)  we propose
a novel abstraction (resp. refinement) strategy to syncretize
abstraction (resp. refinement) steps, achieving significantly better performance
than \cegarnn.

Finally, we remark that any sound tool with the ability to produce counterexamples when properties are violated
could be used as the back-end verification engine in our approach.

\section{Conclusion}\label{sec:conclusion}
We have presented a novel CEGAR-based approach for scalable and exact
verification of neural networks. Specifically, we have proposed two structure-oriented abstraction primitives
and an abstraction strategy to syncretize abstraction
steps, resulting in a novel abstraction procedure.
We also have proposed two corresponding refinement primitives
and a refinement strategy to syncretize refinement
steps, resulting in a novel
refinement procedure. %
We have implemented our approach in a tool
and conducted an extensive evaluation on three widely-used benchmarks.
The results demonstrate that our approach can significantly
boost the scalability and efficiency of two promising and exact verification tools \marabou and \planet without loss of accuracy, in particular, for difficult
verification problems. Our approach also significantly outperforms the unique
structure-oriented CEGAR-based approach \cegarnn, namely, 11.6--26.6 times faster.

\bibliographystyle{ACM-Reference-Format}
\bibliography{main}


\begin{thebibliography}{51}


\ifx \showCODEN    \undefined \def \showCODEN     #1{\unskip}     \fi
\ifx \showDOI      \undefined \def \showDOI       #1{#1}\fi
\ifx \showISBNx    \undefined \def \showISBNx     #1{\unskip}     \fi
\ifx \showISBNxiii \undefined \def \showISBNxiii  #1{\unskip}     \fi
\ifx \showISSN     \undefined \def \showISSN      #1{\unskip}     \fi
\ifx \showLCCN     \undefined \def \showLCCN      #1{\unskip}     \fi
\ifx \shownote     \undefined \def \shownote      #1{#1}          \fi
\ifx \showarticletitle \undefined \def \showarticletitle #1{#1}   \fi
\ifx \showURL      \undefined \def \showURL       {\relax}        \fi
\providecommand\bibfield[2]{#2}
\providecommand\bibinfo[2]{#2}
\providecommand\natexlab[1]{#1}
\providecommand\showeprint[2][]{arXiv:#2}

\bibitem[vnn(2021)]%
        {vnn-comp21}
 \bibinfo{year}{2021}\natexlab{}.
\newblock \bibinfo{booktitle}{\emph{2nd International Verification of Neural
  Networks Competition (VNN-COMP'21)}}.
\newblock
\urldef\tempurl%
\url{https://sites.google.com/view/vnn2021}
\showURL{%
Retrieved 2021 from \tempurl}


\bibitem[Ashok et~al\mbox{.}(2020)]%
        {DBLP:conf/atva/AshokHKM20}
\bibfield{author}{\bibinfo{person}{Pranav Ashok}, \bibinfo{person}{Vahid
  Hashemi}, \bibinfo{person}{Jan Kret{\'{\i}}nsk{\'{y}}}, {and}
  \bibinfo{person}{Stefanie Mohr}.} \bibinfo{year}{2020}\natexlab{}.
\newblock \showarticletitle{DeepAbstract: Neural Network Abstraction for
  Accelerating Verification}. In \bibinfo{booktitle}{\emph{Automated Technology
  for Verification and Analysis - 18th International Symposium, {ATVA} 2020,
  Hanoi, Vietnam, October 19-23, 2020, Proceedings}}
  \emph{(\bibinfo{series}{Lecture Notes in Computer Science},
  Vol.~\bibinfo{volume}{12302})}, \bibfield{editor}{\bibinfo{person}{Dang~Van
  Hung} {and} \bibinfo{person}{Oleg Sokolsky}} (Eds.).
  \bibinfo{publisher}{Springer}, \bibinfo{pages}{92--107}.
\newblock
\urldef\tempurl%
\url{https://doi.org/10.1007/978-3-030-59152-6\_5}
\showDOI{\tempurl}


\bibitem[Bak et~al\mbox{.}(2021)]%
        {abs-2109-00498}
\bibfield{author}{\bibinfo{person}{Stanley Bak}, \bibinfo{person}{Changliu
  Liu}, {and} \bibinfo{person}{Taylor~T. Johnson}.}
  \bibinfo{year}{2021}\natexlab{}.
\newblock \showarticletitle{The Second International Verification of Neural
  Networks Competition {(VNN-COMP} 2021): Summary and Results}.
\newblock \bibinfo{journal}{\emph{CoRR}}  \bibinfo{volume}{abs/2109.00498}
  (\bibinfo{year}{2021}).
\newblock
\showeprint[arXiv]{2109.00498}
\urldef\tempurl%
\url{https://arxiv.org/abs/2109.00498}
\showURL{%
\tempurl}


\bibitem[Bunel et~al\mbox{.}(2017)]%
        {DBLP:journals/corr/abs-1711-00455}
\bibfield{author}{\bibinfo{person}{Rudy Bunel}, \bibinfo{person}{Ilker
  Turkaslan}, \bibinfo{person}{Philip H.~S. Torr}, \bibinfo{person}{Pushmeet
  Kohli}, {and} \bibinfo{person}{M.~Pawan Kumar}.}
  \bibinfo{year}{2017}\natexlab{}.
\newblock \showarticletitle{Piecewise Linear Neural Network verification: {A}
  comparative study}.
\newblock \bibinfo{journal}{\emph{CoRR}}  \bibinfo{volume}{abs/1711.00455}
  (\bibinfo{year}{2017}).
\newblock
\showeprint[arxiv]{1711.00455}
\urldef\tempurl%
\url{http://arxiv.org/abs/1711.00455}
\showURL{%
\tempurl}


\bibitem[Carlini and Wagner(2017)]%
        {Carlini017}
\bibfield{author}{\bibinfo{person}{Nicholas Carlini} {and}
  \bibinfo{person}{David~A. Wagner}.} \bibinfo{year}{2017}\natexlab{}.
\newblock \showarticletitle{Towards Evaluating the Robustness of Neural
  Networks}. In \bibinfo{booktitle}{\emph{2017 {IEEE} Symposium on Security and
  Privacy, {SP} 2017, San Jose, CA, USA, May 22-26, 2017}}.
  \bibinfo{pages}{39--57}.
\newblock
\urldef\tempurl%
\url{https://doi.org/10.1109/SP.2017.49}
\showDOI{\tempurl}


\bibitem[Clarke et~al\mbox{.}(2000)]%
        {DBLP:conf/cav/ClarkeGJLV00}
\bibfield{author}{\bibinfo{person}{Edmund~M. Clarke}, \bibinfo{person}{Orna
  Grumberg}, \bibinfo{person}{Somesh Jha}, \bibinfo{person}{Yuan Lu}, {and}
  \bibinfo{person}{Helmut Veith}.} \bibinfo{year}{2000}\natexlab{}.
\newblock \showarticletitle{Counterexample-Guided Abstraction Refinement}. In
  \bibinfo{booktitle}{\emph{Computer Aided Verification, 12th International
  Conference, {CAV} 2000, Chicago, IL, USA, July 15-19, 2000, Proceedings}}
  \emph{(\bibinfo{series}{Lecture Notes in Computer Science},
  Vol.~\bibinfo{volume}{1855})}, \bibfield{editor}{\bibinfo{person}{E.~Allen
  Emerson} {and} \bibinfo{person}{A.~Prasad Sistla}} (Eds.).
  \bibinfo{publisher}{Springer}, \bibinfo{pages}{154--169}.
\newblock
\urldef\tempurl%
\url{https://doi.org/10.1007/10722167\_15}
\showDOI{\tempurl}


\bibitem[Cousot and Cousot(1977)]%
        {DBLP:conf/popl/CousotC77}
\bibfield{author}{\bibinfo{person}{Patrick Cousot} {and}
  \bibinfo{person}{Radhia Cousot}.} \bibinfo{year}{1977}\natexlab{}.
\newblock \showarticletitle{Abstract Interpretation: {A} Unified Lattice Model
  for Static Analysis of Programs by Construction or Approximation of
  Fixpoints}. In \bibinfo{booktitle}{\emph{Conference Record of the Fourth
  {ACM} Symposium on Principles of Programming Languages, Los Angeles,
  California, USA, January 1977}}, \bibfield{editor}{\bibinfo{person}{Robert~M.
  Graham}, \bibinfo{person}{Michael~A. Harrison}, {and} \bibinfo{person}{Ravi
  Sethi}} (Eds.). \bibinfo{publisher}{{ACM}}, \bibinfo{pages}{238--252}.
\newblock
\urldef\tempurl%
\url{https://doi.org/10.1145/512950.512973}
\showDOI{\tempurl}


\bibitem[Cousot and Halbwachs(1978)]%
        {DBLP:conf/popl/CousotH78}
\bibfield{author}{\bibinfo{person}{Patrick Cousot} {and}
  \bibinfo{person}{Nicolas Halbwachs}.} \bibinfo{year}{1978}\natexlab{}.
\newblock \showarticletitle{Automatic Discovery of Linear Restraints Among
  Variables of a Program}. In \bibinfo{booktitle}{\emph{Conference Record of
  the Fifth Annual {ACM} Symposium on Principles of Programming Languages,
  Tucson, Arizona, USA, January 1978}},
  \bibfield{editor}{\bibinfo{person}{Alfred~V. Aho},
  \bibinfo{person}{Stephen~N. Zilles}, {and} \bibinfo{person}{Thomas~G.
  Szymanski}} (Eds.). \bibinfo{publisher}{{ACM} Press},
  \bibinfo{pages}{84--96}.
\newblock
\urldef\tempurl%
\url{https://doi.org/10.1145/512760.512770}
\showDOI{\tempurl}


\bibitem[Dalvi et~al\mbox{.}(2004)]%
        {DalviDMSV04}
\bibfield{author}{\bibinfo{person}{Nilesh~N. Dalvi}, \bibinfo{person}{Pedro~M.
  Domingos}, \bibinfo{person}{Mausam}, \bibinfo{person}{Sumit~K. Sanghai},
  {and} \bibinfo{person}{Deepak Verma}.} \bibinfo{year}{2004}\natexlab{}.
\newblock \showarticletitle{Adversarial classification}. In
  \bibinfo{booktitle}{\emph{Proceedings of the Tenth {ACM} {SIGKDD}
  International Conference on Knowledge Discovery and Data Mining}}.
  \bibinfo{pages}{99--108}.
\newblock
\urldef\tempurl%
\url{https://doi.org/10.1145/1014052.1014066}
\showDOI{\tempurl}


\bibitem[Dutta et~al\mbox{.}(2018)]%
        {DBLP:conf/nfm/DuttaJST18}
\bibfield{author}{\bibinfo{person}{Souradeep Dutta}, \bibinfo{person}{Susmit
  Jha}, \bibinfo{person}{Sriram Sankaranarayanan}, {and}
  \bibinfo{person}{Ashish Tiwari}.} \bibinfo{year}{2018}\natexlab{}.
\newblock \showarticletitle{Output Range Analysis for Deep Feedforward Neural
  Networks}. In \bibinfo{booktitle}{\emph{{NASA} Formal Methods - 10th
  International Symposium, {NFM} 2018, Newport News, VA, USA, April 17-19,
  2018, Proceedings}} \emph{(\bibinfo{series}{Lecture Notes in Computer
  Science}, Vol.~\bibinfo{volume}{10811})},
  \bibfield{editor}{\bibinfo{person}{Aaron Dutle},
  \bibinfo{person}{C{\'{e}}sar~A. Mu{\~{n}}oz}, {and} \bibinfo{person}{Anthony
  Narkawicz}} (Eds.). \bibinfo{publisher}{Springer}, \bibinfo{pages}{121--138}.
\newblock
\urldef\tempurl%
\url{https://doi.org/10.1007/978-3-319-77935-5\_9}
\showDOI{\tempurl}


\bibitem[Ehlers(2017)]%
        {DBLP:conf/atva/Ehlers17}
\bibfield{author}{\bibinfo{person}{R{\"{u}}diger Ehlers}.}
  \bibinfo{year}{2017}\natexlab{}.
\newblock \showarticletitle{Formal Verification of Piece-Wise Linear
  Feed-Forward Neural Networks}. In \bibinfo{booktitle}{\emph{Automated
  Technology for Verification and Analysis - 15th International Symposium,
  {ATVA} 2017, Pune, India, October 3-6, 2017, Proceedings}}
  \emph{(\bibinfo{series}{Lecture Notes in Computer Science},
  Vol.~\bibinfo{volume}{10482})}, \bibfield{editor}{\bibinfo{person}{Deepak
  D'Souza} {and} \bibinfo{person}{K.~Narayan Kumar}} (Eds.).
  \bibinfo{publisher}{Springer}, \bibinfo{pages}{269--286}.
\newblock
\urldef\tempurl%
\url{https://doi.org/10.1007/978-3-319-68167-2\_19}
\showDOI{\tempurl}


\bibitem[Elboher et~al\mbox{.}(2020)]%
        {DBLP:conf/cav/ElboherGK20}
\bibfield{author}{\bibinfo{person}{Yizhak~Yisrael Elboher},
  \bibinfo{person}{Justin Gottschlich}, {and} \bibinfo{person}{Guy Katz}.}
  \bibinfo{year}{2020}\natexlab{}.
\newblock \showarticletitle{An Abstraction-Based Framework for Neural Network
  Verification}. In \bibinfo{booktitle}{\emph{Computer Aided Verification -
  32nd International Conference, {CAV} 2020, Los Angeles, CA, USA, July 21-24,
  2020, Proceedings, Part {I}}} \emph{(\bibinfo{series}{Lecture Notes in
  Computer Science}, Vol.~\bibinfo{volume}{12224})},
  \bibfield{editor}{\bibinfo{person}{Shuvendu~K. Lahiri} {and}
  \bibinfo{person}{Chao Wang}} (Eds.). \bibinfo{publisher}{Springer},
  \bibinfo{pages}{43--65}.
\newblock
\urldef\tempurl%
\url{https://doi.org/10.1007/978-3-030-53288-8\_3}
\showDOI{\tempurl}


\bibitem[Gehr et~al\mbox{.}(2018)]%
        {DBLP:conf/sp/GehrMDTCV18}
\bibfield{author}{\bibinfo{person}{Timon Gehr}, \bibinfo{person}{Matthew
  Mirman}, \bibinfo{person}{Dana Drachsler{-}Cohen}, \bibinfo{person}{Petar
  Tsankov}, \bibinfo{person}{Swarat Chaudhuri}, {and}
  \bibinfo{person}{Martin~T. Vechev}.} \bibinfo{year}{2018}\natexlab{}.
\newblock \showarticletitle{{AI2:} Safety and Robustness Certification of
  Neural Networks with Abstract Interpretation}. In
  \bibinfo{booktitle}{\emph{2018 {IEEE} Symposium on Security and Privacy, {SP}
  2018, Proceedings, 21-23 May 2018, San Francisco, California, {USA}}}.
  \bibinfo{publisher}{{IEEE} Computer Society}, \bibinfo{pages}{3--18}.
\newblock
\urldef\tempurl%
\url{https://doi.org/10.1109/SP.2018.00058}
\showDOI{\tempurl}


\bibitem[Ghorbal et~al\mbox{.}(2009)]%
        {DBLP:conf/cav/GhorbalGP09}
\bibfield{author}{\bibinfo{person}{Khalil Ghorbal}, \bibinfo{person}{Eric
  Goubault}, {and} \bibinfo{person}{Sylvie Putot}.}
  \bibinfo{year}{2009}\natexlab{}.
\newblock \showarticletitle{The Zonotope Abstract Domain Taylor1+}. In
  \bibinfo{booktitle}{\emph{Computer Aided Verification, 21st International
  Conference, {CAV} 2009, Grenoble, France, June 26 - July 2, 2009.
  Proceedings}} \emph{(\bibinfo{series}{Lecture Notes in Computer Science},
  Vol.~\bibinfo{volume}{5643})}, \bibfield{editor}{\bibinfo{person}{Ahmed
  Bouajjani} {and} \bibinfo{person}{Oded Maler}} (Eds.).
  \bibinfo{publisher}{Springer}, \bibinfo{pages}{627--633}.
\newblock
\urldef\tempurl%
\url{https://doi.org/10.1007/978-3-642-02658-4\_47}
\showDOI{\tempurl}


\bibitem[Gokulanathan et~al\mbox{.}(2020)]%
        {DBLP:conf/nfm/GokulanathanFMB20}
\bibfield{author}{\bibinfo{person}{Sumathi Gokulanathan},
  \bibinfo{person}{Alexander Feldsher}, \bibinfo{person}{Adi Malca},
  \bibinfo{person}{Clark~W. Barrett}, {and} \bibinfo{person}{Guy Katz}.}
  \bibinfo{year}{2020}\natexlab{}.
\newblock \showarticletitle{Simplifying Neural Networks Using Formal
  Verification}. In \bibinfo{booktitle}{\emph{{NASA} Formal Methods - 12th
  International Symposium, {NFM} 2020, Moffett Field, CA, USA, May 11-15, 2020,
  Proceedings}} \emph{(\bibinfo{series}{Lecture Notes in Computer Science},
  Vol.~\bibinfo{volume}{12229})}, \bibfield{editor}{\bibinfo{person}{Ritchie
  Lee}, \bibinfo{person}{Susmit Jha}, {and} \bibinfo{person}{Anastasia
  Mavridou}} (Eds.). \bibinfo{publisher}{Springer}, \bibinfo{pages}{85--93}.
\newblock
\urldef\tempurl%
\url{https://doi.org/10.1007/978-3-030-55754-6\_5}
\showDOI{\tempurl}


\bibitem[Goodfellow et~al\mbox{.}(2015)]%
        {GoodfellowSS14}
\bibfield{author}{\bibinfo{person}{Ian~J. Goodfellow},
  \bibinfo{person}{Jonathon Shlens}, {and} \bibinfo{person}{Christian
  Szegedy}.} \bibinfo{year}{2015}\natexlab{}.
\newblock \showarticletitle{Explaining and Harnessing Adversarial Examples}. In
  \bibinfo{booktitle}{\emph{3rd International Conference on Learning
  Representations, {ICLR} 2015, San Diego, CA, USA, May 7-9, 2015, Conference
  Track Proceedings}}.
\newblock
\urldef\tempurl%
\url{http://arxiv.org/abs/1412.6572}
\showURL{%
\tempurl}


\bibitem[Hinton et~al\mbox{.}(2012)]%
        {DBLP:journals/spm/X12a}
\bibfield{author}{\bibinfo{person}{Geoffrey Hinton}, \bibinfo{person}{Li Deng},
  \bibinfo{person}{Dong Yu}, \bibinfo{person}{George~E. Dahl},
  \bibinfo{person}{Abdel-rahman Mohamed}, \bibinfo{person}{Navdeep Jaitly},
  \bibinfo{person}{Andrew Senior}, \bibinfo{person}{Vincent Vanhoucke},
  \bibinfo{person}{Patrick Nguyen}, \bibinfo{person}{Tara~N. Sainath}, {and}
  \bibinfo{person}{Brian Kingsbury}.} \bibinfo{year}{2012}\natexlab{}.
\newblock \showarticletitle{Deep Neural Networks for Acoustic Modeling in
  Speech Recognition: The Shared Views of Four Research Groups}.
\newblock \bibinfo{journal}{\emph{{IEEE} Signal Process. Mag.}}
  \bibinfo{volume}{29}, \bibinfo{number}{6} (\bibinfo{year}{2012}),
  \bibinfo{pages}{82--97}.
\newblock
\urldef\tempurl%
\url{https://doi.org/10.1109/MSP.2012.2205597}
\showDOI{\tempurl}


\bibitem[Huang et~al\mbox{.}(2017)]%
        {DBLP:conf/cav/HuangKWW17}
\bibfield{author}{\bibinfo{person}{Xiaowei Huang}, \bibinfo{person}{Marta
  Kwiatkowska}, \bibinfo{person}{Sen Wang}, {and} \bibinfo{person}{Min Wu}.}
  \bibinfo{year}{2017}\natexlab{}.
\newblock \showarticletitle{Safety Verification of Deep Neural Networks}. In
  \bibinfo{booktitle}{\emph{Computer Aided Verification - 29th International
  Conference, {CAV} 2017, Heidelberg, Germany, July 24-28, 2017, Proceedings,
  Part {I}}} \emph{(\bibinfo{series}{Lecture Notes in Computer Science},
  Vol.~\bibinfo{volume}{10426})}, \bibfield{editor}{\bibinfo{person}{Rupak
  Majumdar} {and} \bibinfo{person}{Viktor Kuncak}} (Eds.).
  \bibinfo{publisher}{Springer}, \bibinfo{pages}{3--29}.
\newblock
\urldef\tempurl%
\url{https://doi.org/10.1007/978-3-319-63387-9\_1}
\showDOI{\tempurl}


\bibitem[Julian et~al\mbox{.}(2019)]%
        {julian2019deep}
\bibfield{author}{\bibinfo{person}{Kyle~D Julian}, \bibinfo{person}{Mykel~J
  Kochenderfer}, {and} \bibinfo{person}{Michael~P Owen}.}
  \bibinfo{year}{2019}\natexlab{}.
\newblock \showarticletitle{Deep neural network compression for aircraft
  collision avoidance systems}.
\newblock \bibinfo{journal}{\emph{Journal of Guidance, Control, and Dynamics}}
  \bibinfo{volume}{42}, \bibinfo{number}{3} (\bibinfo{year}{2019}),
  \bibinfo{pages}{598--608}.
\newblock


\bibitem[Katz et~al\mbox{.}(2017)]%
        {DBLP:conf/cav/KatzBDJK17}
\bibfield{author}{\bibinfo{person}{Guy Katz}, \bibinfo{person}{Clark~W.
  Barrett}, \bibinfo{person}{David~L. Dill}, \bibinfo{person}{Kyle Julian},
  {and} \bibinfo{person}{Mykel~J. Kochenderfer}.}
  \bibinfo{year}{2017}\natexlab{}.
\newblock \showarticletitle{Reluplex: An Efficient {SMT} Solver for Verifying
  Deep Neural Networks}. In \bibinfo{booktitle}{\emph{Computer Aided
  Verification - 29th International Conference, {CAV} 2017, Heidelberg,
  Germany, July 24-28, 2017, Proceedings, Part {I}}}
  \emph{(\bibinfo{series}{Lecture Notes in Computer Science},
  Vol.~\bibinfo{volume}{10426})}, \bibfield{editor}{\bibinfo{person}{Rupak
  Majumdar} {and} \bibinfo{person}{Viktor Kuncak}} (Eds.).
  \bibinfo{publisher}{Springer}, \bibinfo{pages}{97--117}.
\newblock
\urldef\tempurl%
\url{https://doi.org/10.1007/978-3-319-63387-9\_5}
\showDOI{\tempurl}


\bibitem[Katz et~al\mbox{.}(2019)]%
        {DBLP:conf/cav/KatzHIJLLSTWZDK19}
\bibfield{author}{\bibinfo{person}{Guy Katz}, \bibinfo{person}{Derek~A. Huang},
  \bibinfo{person}{Duligur Ibeling}, \bibinfo{person}{Kyle Julian},
  \bibinfo{person}{Christopher Lazarus}, \bibinfo{person}{Rachel Lim},
  \bibinfo{person}{Parth Shah}, \bibinfo{person}{Shantanu Thakoor},
  \bibinfo{person}{Haoze Wu}, \bibinfo{person}{Aleksandar Zeljic},
  \bibinfo{person}{David~L. Dill}, \bibinfo{person}{Mykel~J. Kochenderfer},
  {and} \bibinfo{person}{Clark~W. Barrett}.} \bibinfo{year}{2019}\natexlab{}.
\newblock \showarticletitle{The Marabou Framework for Verification and Analysis
  of Deep Neural Networks}. In \bibinfo{booktitle}{\emph{Computer Aided
  Verification - 31st International Conference, {CAV} 2019, New York City, NY,
  USA, July 15-18, 2019, Proceedings, Part {I}}}
  \emph{(\bibinfo{series}{Lecture Notes in Computer Science},
  Vol.~\bibinfo{volume}{11561})}, \bibfield{editor}{\bibinfo{person}{Isil
  Dillig} {and} \bibinfo{person}{Serdar Tasiran}} (Eds.).
  \bibinfo{publisher}{Springer}, \bibinfo{pages}{443--452}.
\newblock
\urldef\tempurl%
\url{https://doi.org/10.1007/978-3-030-25540-4\_26}
\showDOI{\tempurl}


\bibitem[Krizhevsky et~al\mbox{.}(2009)]%
        {krizhevsky2009learning}
\bibfield{author}{\bibinfo{person}{Alex Krizhevsky}, \bibinfo{person}{Geoffrey
  Hinton}, {et~al\mbox{.}}} \bibinfo{year}{2009}\natexlab{}.
\newblock \showarticletitle{Learning multiple layers of features from tiny
  images}.
\newblock  (\bibinfo{year}{2009}).
\newblock


\bibitem[Kurakin et~al\mbox{.}(2017)]%
        {KurakinGB17a}
\bibfield{author}{\bibinfo{person}{Alexey Kurakin}, \bibinfo{person}{Ian~J.
  Goodfellow}, {and} \bibinfo{person}{Samy Bengio}.}
  \bibinfo{year}{2017}\natexlab{}.
\newblock \showarticletitle{Adversarial examples in the physical world}. In
  \bibinfo{booktitle}{\emph{5th International Conference on Learning
  Representations, {ICLR} 2017, Toulon, France, April 24-26, 2017, Workshop
  Track Proceedings}}.
\newblock
\urldef\tempurl%
\url{https://openreview.net/forum?id=HJGU3Rodl}
\showURL{%
\tempurl}


\bibitem[Lab(2022)]%
        {eran-repo}
\bibfield{author}{\bibinfo{person}{SRI Lab}.} \bibinfo{year}{2022}\natexlab{}.
\newblock \bibinfo{booktitle}{\emph{ETH Robustness Analyzer for Neural Networks
  (ERAN)}}.
\newblock
\urldef\tempurl%
\url{https://github.com/eth-sri/eran}
\showURL{%
\tempurl}


\bibitem[LeCun(1998)]%
        {lecun1998mnist}
\bibfield{author}{\bibinfo{person}{Yann LeCun}.}
  \bibinfo{year}{1998}\natexlab{}.
\newblock \showarticletitle{The MNIST database of handwritten digits}.
\newblock \bibinfo{journal}{\emph{http://yann. lecun. com/exdb/mnist/}}
  (\bibinfo{year}{1998}).
\newblock


\bibitem[Lin et~al\mbox{.}(2019)]%
        {DBLP:conf/cvpr/LinYCZLLH19}
\bibfield{author}{\bibinfo{person}{Wang Lin}, \bibinfo{person}{Zhengfeng Yang},
  \bibinfo{person}{Xin Chen}, \bibinfo{person}{Qingye Zhao},
  \bibinfo{person}{Xiangkun Li}, \bibinfo{person}{Zhiming Liu}, {and}
  \bibinfo{person}{Jifeng He}.} \bibinfo{year}{2019}\natexlab{}.
\newblock \showarticletitle{Robustness Verification of Classification Deep
  Neural Networks via Linear Programming}. In \bibinfo{booktitle}{\emph{{IEEE}
  Conference on Computer Vision and Pattern Recognition, {CVPR} 2019, Long
  Beach, CA, USA, June 16-20, 2019}}. \bibinfo{publisher}{Computer Vision
  Foundation / {IEEE}}, \bibinfo{pages}{11418--11427}.
\newblock
\urldef\tempurl%
\url{https://doi.org/10.1109/CVPR.2019.01168}
\showDOI{\tempurl}


\bibitem[Litjens et~al\mbox{.}(2017)]%
        {DBLP:journals/mia/LitjensKBSCGLGS17}
\bibfield{author}{\bibinfo{person}{Geert Litjens}, \bibinfo{person}{Thijs
  Kooi}, \bibinfo{person}{Babak~Ehteshami Bejnordi}, \bibinfo{person}{Arnaud
  Arindra~Adiyoso Setio}, \bibinfo{person}{Francesco Ciompi},
  \bibinfo{person}{Mohsen Ghafoorian}, \bibinfo{person}{Jeroen A. W.~M. van~der
  Laak}, \bibinfo{person}{Bram van Ginneken}, {and} \bibinfo{person}{Clara~I.
  S{\'{a}}nchez}.} \bibinfo{year}{2017}\natexlab{}.
\newblock \showarticletitle{A survey on deep learning in medical image
  analysis}.
\newblock \bibinfo{journal}{\emph{Medical Image Anal.}}  \bibinfo{volume}{42}
  (\bibinfo{year}{2017}), \bibinfo{pages}{60--88}.
\newblock
\urldef\tempurl%
\url{https://doi.org/10.1016/j.media.2017.07.005}
\showDOI{\tempurl}


\bibitem[Lomuscio and Maganti(2017)]%
        {DBLP:journals/corr/LomuscioM17}
\bibfield{author}{\bibinfo{person}{Alessio Lomuscio} {and}
  \bibinfo{person}{Lalit Maganti}.} \bibinfo{year}{2017}\natexlab{}.
\newblock \showarticletitle{An approach to reachability analysis for
  feed-forward ReLU neural networks}.
\newblock \bibinfo{journal}{\emph{CoRR}}  \bibinfo{volume}{abs/1706.07351}
  (\bibinfo{year}{2017}).
\newblock
\showeprint[arxiv]{1706.07351}
\urldef\tempurl%
\url{http://arxiv.org/abs/1706.07351}
\showURL{%
\tempurl}


\bibitem[Ma et~al\mbox{.}(2018a)]%
        {MaJZSXLCSLLZW18}
\bibfield{author}{\bibinfo{person}{Lei Ma}, \bibinfo{person}{Felix
  Juefei{-}Xu}, \bibinfo{person}{Fuyuan Zhang}, \bibinfo{person}{Jiyuan Sun},
  \bibinfo{person}{Minhui Xue}, \bibinfo{person}{Bo Li},
  \bibinfo{person}{Chunyang Chen}, \bibinfo{person}{Ting Su},
  \bibinfo{person}{Li Li}, \bibinfo{person}{Yang Liu}, \bibinfo{person}{Jianjun
  Zhao}, {and} \bibinfo{person}{Yadong Wang}.}
  \bibinfo{year}{2018}\natexlab{a}.
\newblock \showarticletitle{DeepGauge: multi-granularity testing criteria for
  deep learning systems}. In \bibinfo{booktitle}{\emph{Proceedings of the 33rd
  {ACM/IEEE} International Conference on Automated Software Engineering}}.
  \bibinfo{pages}{120--131}.
\newblock


\bibitem[Ma et~al\mbox{.}(2018b)]%
        {MaLLZG18}
\bibfield{author}{\bibinfo{person}{Shiqing Ma}, \bibinfo{person}{Yingqi Liu},
  \bibinfo{person}{Wen{-}Chuan Lee}, \bibinfo{person}{Xiangyu Zhang}, {and}
  \bibinfo{person}{Ananth Grama}.} \bibinfo{year}{2018}\natexlab{b}.
\newblock \showarticletitle{{MODE:} automated neural network model debugging
  via state differential analysis and input selection}. In
  \bibinfo{booktitle}{\emph{Proceedings of the 2018 {ACM} Joint Meeting on
  European Software Engineering Conference and Symposium on the Foundations of
  Software Engineering}}. \bibinfo{pages}{175--186}.
\newblock


\bibitem[Ma and Lu(2017)]%
        {abs-1712-01252}
\bibfield{author}{\bibinfo{person}{Wei Ma} {and} \bibinfo{person}{Jun Lu}.}
  \bibinfo{year}{2017}\natexlab{}.
\newblock \showarticletitle{An Equivalence of Fully Connected Layer and
  Convolutional Layer}.
\newblock \bibinfo{journal}{\emph{CoRR}}  \bibinfo{volume}{abs/1712.01252}
  (\bibinfo{year}{2017}).
\newblock
\showeprint[arXiv]{1712.01252}
\urldef\tempurl%
\url{http://arxiv.org/abs/1712.01252}
\showURL{%
\tempurl}


\bibitem[Ostrovsky et~al\mbox{.}(2022)]%
        {DBLP:journals/corr/abs-2201-01978}
\bibfield{author}{\bibinfo{person}{Matan Ostrovsky}, \bibinfo{person}{Clark~W.
  Barrett}, {and} \bibinfo{person}{Guy Katz}.} \bibinfo{year}{2022}\natexlab{}.
\newblock \showarticletitle{An Abstraction-Refinement Approach to Verifying
  Convolutional Neural Networks}.
\newblock \bibinfo{journal}{\emph{CoRR}}  \bibinfo{volume}{abs/2201.01978}
  (\bibinfo{year}{2022}).
\newblock
\showeprint[arXiv]{2201.01978}
\urldef\tempurl%
\url{https://arxiv.org/abs/2201.01978}
\showURL{%
\tempurl}


\bibitem[Papernot et~al\mbox{.}(2016)]%
        {PapernotMJFCS16}
\bibfield{author}{\bibinfo{person}{Nicolas Papernot},
  \bibinfo{person}{Patrick~D. McDaniel}, \bibinfo{person}{Somesh Jha},
  \bibinfo{person}{Matt Fredrikson}, \bibinfo{person}{Z.~Berkay Celik}, {and}
  \bibinfo{person}{Ananthram Swami}.} \bibinfo{year}{2016}\natexlab{}.
\newblock \showarticletitle{The Limitations of Deep Learning in Adversarial
  Settings}. In \bibinfo{booktitle}{\emph{{IEEE} European Symposium on Security
  and Privacy, EuroS{\&}P 2016, Saarbr{\"{u}}cken, Germany, March 21-24,
  2016}}. \bibinfo{pages}{372--387}.
\newblock
\urldef\tempurl%
\url{https://doi.org/10.1109/EuroSP.2016.36}
\showDOI{\tempurl}


\bibitem[Paulsen and Wang(2022)]%
        {PW22}
\bibfield{author}{\bibinfo{person}{Brandon Paulsen} {and} \bibinfo{person}{Chao
  Wang}.} \bibinfo{year}{2022}\natexlab{}.
\newblock \showarticletitle{LinSyn: Synthesizing Tight Linear Bounds for
  Arbitrary Neural Network Activation Functions}.
\newblock \bibinfo{journal}{\emph{CoRR}}  \bibinfo{volume}{abs/2201.13351}
  (\bibinfo{year}{2022}).
\newblock


\bibitem[Pei et~al\mbox{.}(2019)]%
        {PeiCYJ19}
\bibfield{author}{\bibinfo{person}{Kexin Pei}, \bibinfo{person}{Yinzhi Cao},
  \bibinfo{person}{Junfeng Yang}, {and} \bibinfo{person}{Suman Jana}.}
  \bibinfo{year}{2019}\natexlab{}.
\newblock \showarticletitle{DeepXplore: automated whitebox testing of deep
  learning systems}.
\newblock \bibinfo{journal}{\emph{Commun. {ACM}}} \bibinfo{volume}{62},
  \bibinfo{number}{11} (\bibinfo{year}{2019}), \bibinfo{pages}{137--145}.
\newblock


\bibitem[Prabhakar and Afzal(2019)]%
        {DBLP:conf/nips/PrabhakarA19}
\bibfield{author}{\bibinfo{person}{Pavithra Prabhakar} {and}
  \bibinfo{person}{Zahra~Rahimi Afzal}.} \bibinfo{year}{2019}\natexlab{}.
\newblock \showarticletitle{Abstraction based Output Range Analysis for Neural
  Networks}. In \bibinfo{booktitle}{\emph{Advances in Neural Information
  Processing Systems 32: Annual Conference on Neural Information Processing
  Systems 2019, NeurIPS 2019, December 8-14, 2019, Vancouver, BC, Canada}},
  \bibfield{editor}{\bibinfo{person}{Hanna~M. Wallach}, \bibinfo{person}{Hugo
  Larochelle}, \bibinfo{person}{Alina Beygelzimer}, \bibinfo{person}{Florence
  d'Alch{\'{e}}{-}Buc}, \bibinfo{person}{Emily~B. Fox}, {and}
  \bibinfo{person}{Roman Garnett}} (Eds.). \bibinfo{pages}{15762--15772}.
\newblock


\bibitem[Pulina and Tacchella(2010)]%
        {DBLP:conf/cav/PulinaT10}
\bibfield{author}{\bibinfo{person}{Luca Pulina} {and} \bibinfo{person}{Armando
  Tacchella}.} \bibinfo{year}{2010}\natexlab{}.
\newblock \showarticletitle{An Abstraction-Refinement Approach to Verification
  of Artificial Neural Networks}. In \bibinfo{booktitle}{\emph{Computer Aided
  Verification, 22nd International Conference, {CAV} 2010, Edinburgh, UK, July
  15-19, 2010. Proceedings}} \emph{(\bibinfo{series}{Lecture Notes in Computer
  Science}, Vol.~\bibinfo{volume}{6174})},
  \bibfield{editor}{\bibinfo{person}{Tayssir Touili}, \bibinfo{person}{Byron
  Cook}, {and} \bibinfo{person}{Paul~B. Jackson}} (Eds.).
  \bibinfo{publisher}{Springer}, \bibinfo{pages}{243--257}.
\newblock
\urldef\tempurl%
\url{https://doi.org/10.1007/978-3-642-14295-6\_24}
\showDOI{\tempurl}


\bibitem[Russakovsky et~al\mbox{.}(2015)]%
        {DBLP:journals/ijcv/RussakovskyDSKS15}
\bibfield{author}{\bibinfo{person}{Olga Russakovsky}, \bibinfo{person}{Jia
  Deng}, \bibinfo{person}{Hao Su}, \bibinfo{person}{Jonathan Krause},
  \bibinfo{person}{Sanjeev Satheesh}, \bibinfo{person}{Sean Ma},
  \bibinfo{person}{Zhiheng Huang}, \bibinfo{person}{Andrej Karpathy},
  \bibinfo{person}{Aditya Khosla}, \bibinfo{person}{Michael~S. Bernstein},
  \bibinfo{person}{Alexander~C. Berg}, {and} \bibinfo{person}{Fei{-}Fei Li}.}
  \bibinfo{year}{2015}\natexlab{}.
\newblock \showarticletitle{ImageNet Large Scale Visual Recognition Challenge}.
\newblock \bibinfo{journal}{\emph{Int. J. Comput. Vis.}} \bibinfo{volume}{115},
  \bibinfo{number}{3} (\bibinfo{year}{2015}), \bibinfo{pages}{211--252}.
\newblock
\urldef\tempurl%
\url{https://doi.org/10.1007/s11263-015-0816-y}
\showDOI{\tempurl}


\bibitem[Singh et~al\mbox{.}(2018)]%
        {DBLP:conf/nips/SinghGMPV18}
\bibfield{author}{\bibinfo{person}{Gagandeep Singh}, \bibinfo{person}{Timon
  Gehr}, \bibinfo{person}{Matthew Mirman}, \bibinfo{person}{Markus
  P{\"{u}}schel}, {and} \bibinfo{person}{Martin~T. Vechev}.}
  \bibinfo{year}{2018}\natexlab{}.
\newblock \showarticletitle{Fast and Effective Robustness Certification}. In
  \bibinfo{booktitle}{\emph{Advances in Neural Information Processing Systems
  31: Annual Conference on Neural Information Processing Systems 2018, NeurIPS
  2018, December 3-8, 2018, Montr{\'{e}}al, Canada}},
  \bibfield{editor}{\bibinfo{person}{Samy Bengio}, \bibinfo{person}{Hanna~M.
  Wallach}, \bibinfo{person}{Hugo Larochelle}, \bibinfo{person}{Kristen
  Grauman}, \bibinfo{person}{Nicol{\`{o}} Cesa{-}Bianchi}, {and}
  \bibinfo{person}{Roman Garnett}} (Eds.). \bibinfo{pages}{10825--10836}.
\newblock


\bibitem[Singh et~al\mbox{.}(2019)]%
        {DBLP:journals/pacmpl/SinghGPV19}
\bibfield{author}{\bibinfo{person}{Gagandeep Singh}, \bibinfo{person}{Timon
  Gehr}, \bibinfo{person}{Markus P{\"{u}}schel}, {and}
  \bibinfo{person}{Martin~T. Vechev}.} \bibinfo{year}{2019}\natexlab{}.
\newblock \showarticletitle{An abstract domain for certifying neural networks}.
\newblock \bibinfo{journal}{\emph{Proc. {ACM} Program. Lang.}}
  \bibinfo{volume}{3}, \bibinfo{number}{{POPL}} (\bibinfo{year}{2019}),
  \bibinfo{pages}{41:1--41:30}.
\newblock
\urldef\tempurl%
\url{https://doi.org/10.1145/3290354}
\showDOI{\tempurl}


\bibitem[Sun et~al\mbox{.}(2018)]%
        {SunWRHKK18}
\bibfield{author}{\bibinfo{person}{Youcheng Sun}, \bibinfo{person}{Min Wu},
  \bibinfo{person}{Wenjie Ruan}, \bibinfo{person}{Xiaowei Huang},
  \bibinfo{person}{Marta Kwiatkowska}, {and} \bibinfo{person}{Daniel
  Kroening}.} \bibinfo{year}{2018}\natexlab{}.
\newblock \showarticletitle{Concolic testing for deep neural networks}. In
  \bibinfo{booktitle}{\emph{Proceedings of the 33rd {ACM/IEEE} International
  Conference on Automated Software Engineering}}. \bibinfo{pages}{109--119}.
\newblock


\bibitem[Szegedy et~al\mbox{.}(2014)]%
        {SzegedyZSBEGF13}
\bibfield{author}{\bibinfo{person}{Christian Szegedy},
  \bibinfo{person}{Wojciech Zaremba}, \bibinfo{person}{Ilya Sutskever},
  \bibinfo{person}{Joan Bruna}, \bibinfo{person}{Dumitru Erhan},
  \bibinfo{person}{Ian~J. Goodfellow}, {and} \bibinfo{person}{Rob Fergus}.}
  \bibinfo{year}{2014}\natexlab{}.
\newblock \showarticletitle{Intriguing properties of neural networks}. In
  \bibinfo{booktitle}{\emph{Proceedings of the 2nd International Conference on
  Learning Representations}}.
\newblock


\bibitem[Tian et~al\mbox{.}(2018)]%
        {TianPJR18}
\bibfield{author}{\bibinfo{person}{Yuchi Tian}, \bibinfo{person}{Kexin Pei},
  \bibinfo{person}{Suman Jana}, {and} \bibinfo{person}{Baishakhi Ray}.}
  \bibinfo{year}{2018}\natexlab{}.
\newblock \showarticletitle{DeepTest: automated testing of
  deep-neural-network-driven autonomous cars}. In
  \bibinfo{booktitle}{\emph{Proceedings of the 40th International Conference on
  Software Engineering}}. \bibinfo{pages}{303--314}.
\newblock


\bibitem[Tjeng and Tedrake(2017)]%
        {DBLP:journals/corr/abs-1711-07356}
\bibfield{author}{\bibinfo{person}{Vincent Tjeng} {and} \bibinfo{person}{Russ
  Tedrake}.} \bibinfo{year}{2017}\natexlab{}.
\newblock \showarticletitle{Verifying Neural Networks with Mixed Integer
  Programming}.
\newblock \bibinfo{journal}{\emph{CoRR}}  \bibinfo{volume}{abs/1711.07356}
  (\bibinfo{year}{2017}).
\newblock
\showeprint[arxiv]{1711.07356}
\urldef\tempurl%
\url{http://arxiv.org/abs/1711.07356}
\showURL{%
\tempurl}


\bibitem[Tran et~al\mbox{.}(2019)]%
        {DBLP:conf/fm/TranLMYNXJ19}
\bibfield{author}{\bibinfo{person}{Hoang{-}Dung Tran},
  \bibinfo{person}{Diego~Manzanas Lopez}, \bibinfo{person}{Patrick Musau},
  \bibinfo{person}{Xiaodong Yang}, \bibinfo{person}{Luan~Viet Nguyen},
  \bibinfo{person}{Weiming Xiang}, {and} \bibinfo{person}{Taylor~T. Johnson}.}
  \bibinfo{year}{2019}\natexlab{}.
\newblock \showarticletitle{Star-Based Reachability Analysis of Deep Neural
  Networks}. In \bibinfo{booktitle}{\emph{Formal Methods - The Next 30 Years -
  Third World Congress, {FM} 2019, Porto, Portugal, October 7-11, 2019,
  Proceedings}} \emph{(\bibinfo{series}{Lecture Notes in Computer Science},
  Vol.~\bibinfo{volume}{11800})}, \bibfield{editor}{\bibinfo{person}{Maurice~H.
  ter Beek}, \bibinfo{person}{Annabelle McIver}, {and}
  \bibinfo{person}{Jos{\'{e}}~N. Oliveira}} (Eds.).
  \bibinfo{publisher}{Springer}, \bibinfo{pages}{670--686}.
\newblock
\urldef\tempurl%
\url{https://doi.org/10.1007/978-3-030-30942-8\_39}
\showDOI{\tempurl}


\bibitem[Urban and Min{\'e}(2021)]%
        {urban2021review}
\bibfield{author}{\bibinfo{person}{Caterina Urban} {and}
  \bibinfo{person}{Antoine Min{\'e}}.} \bibinfo{year}{2021}\natexlab{}.
\newblock \showarticletitle{A Review of Formal Methods applied to Machine
  Learning}.
\newblock \bibinfo{journal}{\emph{arXiv preprint arXiv:2104.02466}}
  (\bibinfo{year}{2021}).
\newblock


\bibitem[Urmson and Whittaker(2008)]%
        {DBLP:journals/expert/UrmsonW08}
\bibfield{author}{\bibinfo{person}{Chris Urmson} {and} \bibinfo{person}{William
  Whittaker}.} \bibinfo{year}{2008}\natexlab{}.
\newblock \showarticletitle{Self-Driving Cars and the Urban Challenge}.
\newblock \bibinfo{journal}{\emph{{IEEE} Intell. Syst.}} \bibinfo{volume}{23},
  \bibinfo{number}{2} (\bibinfo{year}{2008}), \bibinfo{pages}{66--68}.
\newblock
\urldef\tempurl%
\url{https://doi.org/10.1109/MIS.2008.34}
\showDOI{\tempurl}


\bibitem[Wang et~al\mbox{.}(2018)]%
        {DBLP:conf/uss/WangPWYJ18}
\bibfield{author}{\bibinfo{person}{Shiqi Wang}, \bibinfo{person}{Kexin Pei},
  \bibinfo{person}{Justin Whitehouse}, \bibinfo{person}{Junfeng Yang}, {and}
  \bibinfo{person}{Suman Jana}.} \bibinfo{year}{2018}\natexlab{}.
\newblock \showarticletitle{Formal Security Analysis of Neural Networks using
  Symbolic Intervals}. In \bibinfo{booktitle}{\emph{27th {USENIX} Security
  Symposium, {USENIX} Security 2018, Baltimore, MD, USA, August 15-17, 2018}},
  \bibfield{editor}{\bibinfo{person}{William Enck} {and}
  \bibinfo{person}{Adrienne~Porter Felt}} (Eds.). \bibinfo{publisher}{{USENIX}
  Association}, \bibinfo{pages}{1599--1614}.
\newblock
\urldef\tempurl%
\url{https://www.usenix.org/conference/usenixsecurity18/presentation/wang-shiqi}
\showURL{%
\tempurl}


\bibitem[Wong and Kolter(2018)]%
        {DBLP:conf/icml/WongK18}
\bibfield{author}{\bibinfo{person}{Eric Wong} {and} \bibinfo{person}{J.~Zico
  Kolter}.} \bibinfo{year}{2018}\natexlab{}.
\newblock \showarticletitle{Provable Defenses against Adversarial Examples via
  the Convex Outer Adversarial Polytope}. In
  \bibinfo{booktitle}{\emph{Proceedings of the 35th International Conference on
  Machine Learning, {ICML} 2018, Stockholmsm{\"{a}}ssan, Stockholm, Sweden,
  July 10-15, 2018}} \emph{(\bibinfo{series}{Proceedings of Machine Learning
  Research}, Vol.~\bibinfo{volume}{80})},
  \bibfield{editor}{\bibinfo{person}{Jennifer~G. Dy} {and}
  \bibinfo{person}{Andreas Krause}} (Eds.). \bibinfo{publisher}{{PMLR}},
  \bibinfo{pages}{5283--5292}.
\newblock
\urldef\tempurl%
\url{http://proceedings.mlr.press/v80/wong18a.html}
\showURL{%
\tempurl}


\bibitem[Yang et~al\mbox{.}(2021)]%
        {DBLP:conf/tacas/YangLLHWSXZ21}
\bibfield{author}{\bibinfo{person}{Pengfei Yang}, \bibinfo{person}{Renjue Li},
  \bibinfo{person}{Jianlin Li}, \bibinfo{person}{Cheng{-}Chao Huang},
  \bibinfo{person}{Jingyi Wang}, \bibinfo{person}{Jun Sun},
  \bibinfo{person}{Bai Xue}, {and} \bibinfo{person}{Lijun Zhang}.}
  \bibinfo{year}{2021}\natexlab{}.
\newblock \showarticletitle{Improving Neural Network Verification through
  Spurious Region Guided Refinement}. In \bibinfo{booktitle}{\emph{Tools and
  Algorithms for the Construction and Analysis of Systems - 27th International
  Conference, {TACAS} 2021, Held as Part of the European Joint Conferences on
  Theory and Practice of Software, {ETAPS} 2021, Luxembourg City, Luxembourg,
  March 27 - April 1, 2021, Proceedings, Part {I}}}
  \emph{(\bibinfo{series}{Lecture Notes in Computer Science},
  Vol.~\bibinfo{volume}{12651})}, \bibfield{editor}{\bibinfo{person}{Jan~Friso
  Groote} {and} \bibinfo{person}{Kim~Guldstrand Larsen}} (Eds.).
  \bibinfo{publisher}{Springer}, \bibinfo{pages}{389--408}.
\newblock
\urldef\tempurl%
\url{https://doi.org/10.1007/978-3-030-72016-2\_21}
\showDOI{\tempurl}


\bibitem[Zhang et~al\mbox{.}(2022)]%
        {ZhangHML22}
\bibfield{author}{\bibinfo{person}{Jie~M. Zhang}, \bibinfo{person}{Mark
  Harman}, \bibinfo{person}{Lei Ma}, {and} \bibinfo{person}{Yang Liu}.}
  \bibinfo{year}{2022}\natexlab{}.
\newblock \showarticletitle{Machine Learning Testing: Survey, Landscapes and
  Horizons}.
\newblock \bibinfo{journal}{\emph{{IEEE} Trans. Software Eng.}}
  \bibinfo{volume}{48}, \bibinfo{number}{2} (\bibinfo{year}{2022}),
  \bibinfo{pages}{1--36}.
\newblock


\end{thebibliography}

\end{document}